\documentclass[a4paper, 11pt]{article}
\usepackage{comment} 
\usepackage{lipsum} 
\usepackage{fullpage} 
\usepackage{amsmath,mathtools,bm,amssymb}
\usepackage{enumitem}
\usepackage{xspace}
\usepackage{pifont}
\usepackage{dsfont}
\usepackage{algpseudocode,algorithm}
\usepackage{graphicx}
\usepackage{subfig}
\usepackage{grffile}
\usepackage{pdflscape}
\usepackage{breqn}
\usepackage{amsthm}
\usepackage{color}
\usepackage{url}
\allowdisplaybreaks
\graphicspath{{./figs/}}

\newtheorem{proposition}{Proposition}[section]
\newtheorem{theorem}{Theorem}[section]

\newtheorem{claim}{Claim}[section]
\newtheorem{definition}{Definition}[section]
\def\done{\hspace*{\fill} \rule{1.8mm}{2.5mm}}
\newcommand{\rmax}{R_{\max}\xspace}

\newcommand{\wmax}{W_{\max}\xspace}
\newcommand{\wtau}{W_{\tau}\xspace}
\newcommand{\wloss}{W_{loss}\xspace}
\newcommand{\rc}{R_{\mathrm{c}}}

\newcommand{\cwnd}{\emph{cwnd}\xspace}

\newcommand{\ptau}{p_{\tau}\xspace}
\algnewcommand{\LineComment}[1]{\State \(\triangleright\) #1}

\newcommand{\etal}{\emph{et al.}\xspace}

\newcommand{\wmaxdot}{\dot{W}_{\max}\xspace}
\newcommand{\wmaxhat}{\hat{W}_{\max}\xspace}
\newcommand{\vect}[1]{\mathbf{#1}}
\newcommand{\psitau}{\Psi_{\tau}}
\newcommand{\ptildetau}{\tilde{p}_{\tau}}
\newcommand{\what}{\hat{W}}
\newcommand{\xotau}{x_{1_{\tau}}}
\newcommand{\xttau}{x_{2_{\tau}}}
\newcommand{\xotheta}{x_{1_{\theta}}}
\newcommand{\xttheta}{x_{2_{\theta}}}

\newcommand{\shat}{\hat{s}}
\newcommand{\phat}{\hat{p}}
\newcommand{\xtdot}{\dot{x}_2}
\newcommand{\xodot}{\dot{x}_1}

\author{
  Gayane Vardoyan$^{*}$\\
  \texttt{gvardoyan@cs.umass.edu}\\
  \and
  C.V. Hollot$^{\dagger}$\\
  \texttt{hollot@ecs.umass.edu}
  \and
  Don Towsley$^{*}$\\
  \texttt{towsley@cs.umass.edu}
}

\begin{document}
\title{Towards Stability Analysis of Data Transport Mechanisms: a Fluid Model and an Application}
\date{\vspace{-0.5em}\small{$^{*}$College of Information and Computer Sciences, $^{\dagger}$Department of Electrical and Computer Engineering
University of Massachusetts, Amherst}}
\maketitle
\begin{abstract}
The Transmission Control Protocol (TCP) utilizes congestion avoidance and control mechanisms as a preventive measure against congestive collapse and as an adaptive measure in the presence of changing network conditions. The set of available congestion control algorithms is diverse, and while many have been studied from empirical and simulation perspectives, there is a notable lack of analytical work for some variants. To gain more insight into the dynamics of these algorithms, we: (1) propose a general modeling scheme consisting of a set of functional differential equations of retarded type (RFDEs) and of the congestion window as a function of time; (2) apply this scheme to TCP Reno and demonstrate its equivalence to a previous, well known model for TCP Reno; (3) show an application of the new framework to the widely-deployed congestion control algorithm TCP CUBIC, for which analytical models are few and limited; 
and (4) validate the model using simulations. Our modeling framework yields 
a fluid model for TCP CUBIC.
From a theoretical analysis of this model, we discover that TCP CUBIC is \textit{locally uniformly asymptotically stable} -- a property of the algorithm previously unknown.
\end{abstract}
\section{Introduction}
\label{intro}
TCP carries most of the traffic on the Internet. One of its important functions is to perform end-to-end congestion control to alleviate congestion in the Internet and to provide fair bandwidth sharing among different flows.  To date, many different congestion control algorithms (variants) have been developed, among which are Reno, Vegas, STCP \cite{kelly2003scalable}, CUBIC \cite{ha2008cubic}, H-TCP \cite{htcp}, and BBR \cite{cardwell2016bbr}. 
Stability is an imperative property for any dynamical system.
The stability of several of these variants including Reno, Vegas, and STCP has been extensively and carefully studied, however, little is known about the stability properties of more recent variants such as CUBIC and {H-TCP}. These latter variants have typically been studied through simulation and experimentation, neither of which are adequate to make careful statements about stability.  
As we will observe, for some variants this deficiency is due to the lack of a modeling framework with which to develop appropriate models that are amenable to a formal stability analysis.
The goals of this paper are to point out deficiencies in the previous framework used to study variants such as Reno that make it unsuitable to study a variant such as CUBIC, and then to present a new framework and apply it to the analysis of CUBIC.  Our choice of CUBIC is because it is a popular variant that is the default in the Linux distribution.

The traditional approach for modeling a congestion control algorithm's behavior is to derive a differential equation (DE) for its congestion window (\cwnd) or sending rate as a function of time. Such DEs typically include the algorithm's increase and decrease rules, as well as loss probability functions, for example, to incorporate an active queue management (AQM) policy. This method is highly effective for modeling certain types of controllers, such as TCP Reno and STCP, whose \cwnd update rules are very simple (\emph{e.g.} Reno's \cwnd grows by one every round trip and decreases by half upon congestion detection). However, this approach reaches its limitations when presented with a controller whose \cwnd update functions are complex, thereby making it difficult or impossible to write a DE for the \cwnd or sending rate directly. For example, CUBIC's increase update rule is a function of \emph{time since last loss} and of the \emph{congestion window size immediately before loss}. Moreover, in the case of CUBIC, the steady-state value of \cwnd lies at the saddle point of the window function, which obstructs the stability analysis of this point of interest.

To overcome the impediments of the traditional approach, we develop a novel framework that exploits the fact that all \cwnd- and rate-based controllers
that utilize packet loss information\footnote[1]{Note that this includes not only ACK-based algorithms, but also packet marking schemes as in ECN (Explicit Congestion Notification). From this point forward, we refer to such schemes as ``loss-based''.} to make changes to the \cwnd or rate
have two variables in common: the value of \cwnd (rate) immediately before loss and the time elapsed since last loss. As a consequence, one can derive a set of two DEs: the first for describing the maximum \cwnd (rate) as a function of time, and the second for describing the duration of congestion epochs. This is a relatively easy task, compared to deriving a DE for \cwnd (rate) of a complex algorithm directly.
The advantage of such a model is that it offers tremendous versatility since
\if{false}:
\begin{enumerate}[label=(\alph*),topsep=0pt,itemsep=-1ex,partopsep=1ex,parsep=1ex]
\item it decouples \cwnd or rate functions from the two DEs, with the latter being identical for many controllers,
\item it can accommodate a diverse set of loss functions $p(t)$, which can include capacity and buffer constraints or different queue policies,
\item it can capture the transient and steady-state behavior of multiple parallel flows.
 \end{enumerate}
 \fi
 it does not define \cwnd or rate functions within the set of DEs, with the latter being identical for many controllers.
 Note that the proposed model is applicable not only to TCP-based congestion controllers, but also to UDT \cite{gu2007udt} and QCN \cite{pan2007qcn}.

In this work, we use simulation to validate our analytical models. As NS3 \cite{ns3} does not natively support CUBIC, and existing implementations of this protocol have known problems,
we introduce 
a lightweight simulation framework that is easily programmed to switch between different congestion control variants.
We use this framework to validate the DE model and observe that the average \cwnd predicted by both are in close agreement. 
As system parameters are varied, the simulation
and the DE model agree on whether the system is stable. For TCP CUBIC, we observe that instability can be introduced by deviating the initial conditions too far from their fixed-point values. While our analysis states that CUBIC is locally asymptotically stable, these simulations complement the theory by demonstrating that CUBIC is \emph{not} globally stable.

The contributions of this work are as follows:
\begin{itemize}[topsep=0pt,itemsep=-1ex,partopsep=1ex,parsep=1ex]
\item a new modeling framework applicable to a diverse set of algorithms,
\item an application of this model to CUBIC, and a stability analysis of this algorithm,
\item validation of this model with simulation; the simulation is of independent interest separate from this paper (a description of the framework is provided in the Appendix).
\end{itemize}

The rest of this paper is organized as follows: we discuss related work in Section \ref{relatedwork}.
We introduce the modeling framework in Section \ref{newmodel} and apply it to TCP Reno. 
In Section \ref{reno}, we show that the new framework is equivalent to the one presented in \cite{Misra:2000:FAN:347057.347421}. 
In Section \ref{cubicmodel}, we apply the new modeling scheme to TCP CUBIC. In the remainder of Section \ref{cubic}, we perform a careful stability analysis of CUBIC and present a convergence result. In Section \ref{simulations}, we validate the new model and the stability result for CUBIC using simulations.
We draw conclusions in Section \ref{conclusion}.
\section{Background}
\label{relatedwork}
There exist a number of analytical studies for modeling TCP and analyzing its stability. In  \cite{Misra:2000:FAN:347057.347421}, Misra \etal derive a fluid model for a set of TCP Reno flows and show an application to a networked setting where RED (Random Early Detection) is the AQM policy.
Kelly proposed an optimization-based framework for studying and designing congestion control algorithms in \cite{kelly1998rate}, where STCP was an output.
In \cite{srikant2012mathematics}, Srikant presented a simple analysis of Jacobson's TCP congestion control algorithm.
In \cite{hollot2001control}, Hollot \etal analyze the stability of TCP with 
an AQM
system implementing RED.

Huang \etal develop and analyze the stability of a general nonlinear model of TCP in \cite{huang2006generalized}, focusing on HighSpeed, Scalable, and Standard TCP for comparisons of relative stability.
The authors rely on functions $f(w)$ and $g(w)$, which are additive and multiplicative parameters, respectively, and are both functions of the current congestion window size.
Our model contrasts from these examples in that rather than modeling the congestion window directly, we instead model two interdependent variables (maximum \cwnd and time between losses) that in turn determine the evolution of the window. This new method presents a window of opportunity for modeling complex, nonlinear transport algorithms for which it is not possible to write a DE for \cwnd directly or whose $f(w)$ and $g(w)$ functions cannot be written in closed form.

Bao \emph{et al.} propose Markov chain models for average steady-state TCP CUBIC throughput, in a wireless environment \cite{5684172}. In \cite{poojary2015asymptotic}, Poojary \emph{et al.} derive an expression for average \cwnd of a single CUBIC flow under random losses. In contrast to \cite{5684172} and \cite{poojary2015asymptotic}, the model we present in this work for CUBIC provides insight into both the transient and steady-state behavior of the algorithm. Moreover, we utilize Lyapunov stability theory to prove that CUBIC is locally asymptotically stable independent of link delay and other system parameters (the parameters only affect the region of stability). This result is one of the main contributions of this work.
\section{The Model}
\label{model}
In this section, we present the new model, which is the focus of this work. As a proof of concept, we apply this model to TCP Reno and show that it is mathematically equivalent to the well-known DE model originally presented in \cite{Misra:2000:FAN:347057.347421}. We note that while the two models are equivalent, they make use of different types of information, which is essential for developing a fluid model for TCP CUBIC presented in Section \ref{cubic}.

In the analysis that follows, we will use the notation $f\equiv f(t)$ to represent a function or variable that is not time-delayed. Similarly, we will use $f_{T}\equiv f(t-T)$ to represent a function or variable that is delayed by an amount of time $T$. We will also use $\dot{f}=df(t)/dt$ to represent a function or variable $f$ differentiated with respect to time.
\subsection{The New Model}
\label{newmodel}
\begin{figure}
\centering
\includegraphics[width=0.5\textwidth]{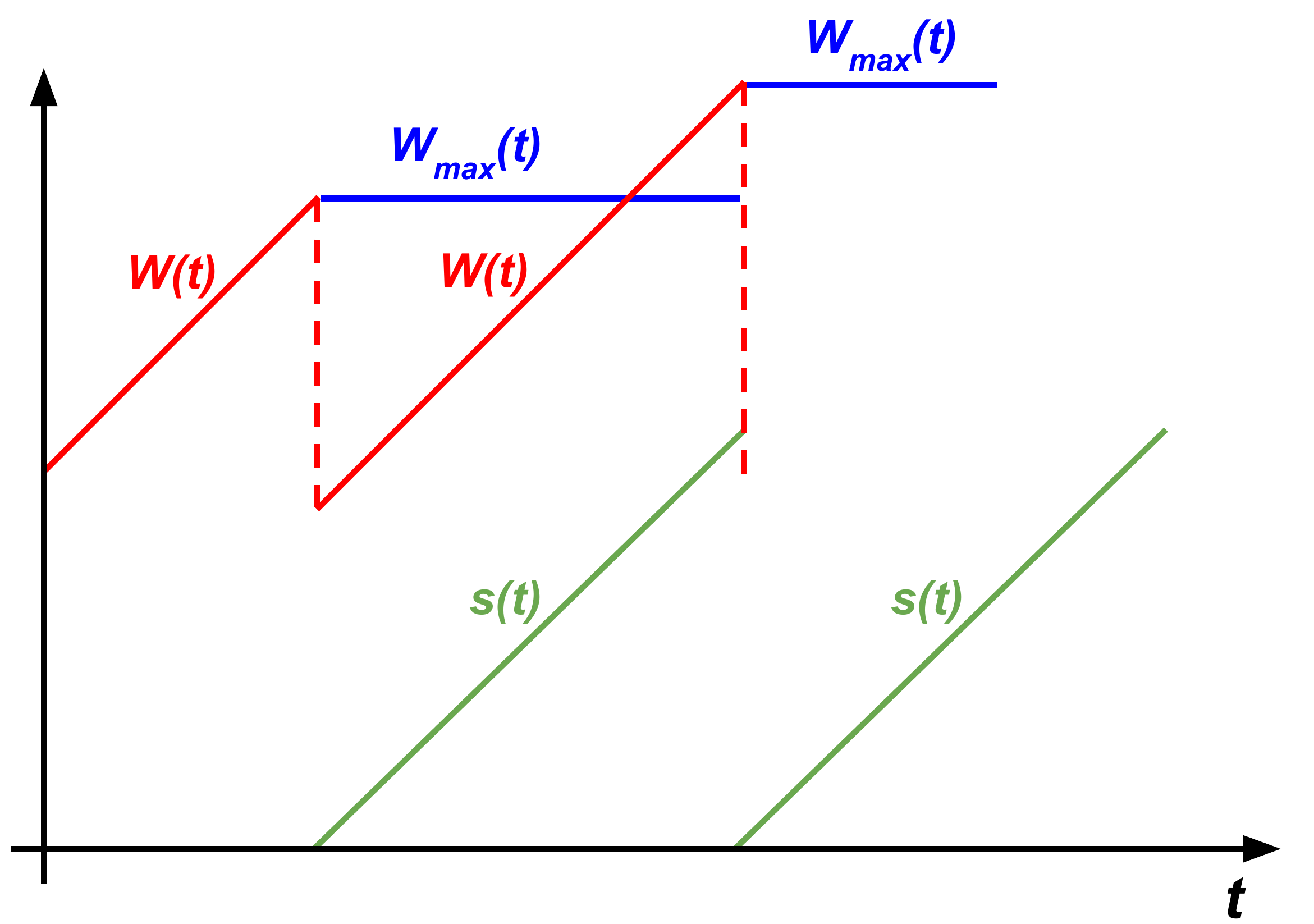}
\caption{$W(t)$, $\wmax(t)$, and $s(t)$ for TCP Reno.}
\label{fig:renofig}
\end{figure}
\begin{table}
\centering
\begin{tabular}{ l | l }
Term & Definition\\\hline
$C$ & per-flow capacity\\
$\tau$ & link delay\\
$\wmax(t)$ & the size of the \cwnd immediately before loss\\
$s(t)$ & the time elapsed since loss\\
$W(t)$ & the \cwnd as a function of time\\
$p(t)$ & a probability of loss function
\end{tabular}
\caption{Term definitions.}
\label{tab:defs}
\end{table}
Table \ref{tab:defs} presents some useful definitions.
The main idea behind the model is the following:
instead of deriving a DE for the \cwnd function $W(t)$ directly, which is specific to a data transport algorithm, we instead derive DEs for $\wmax(t)$ -- the size of the \cwnd immediately before loss, and $s(t)$ -- the amount of time elapsed since last loss, which are variables common to all loss-based algorithms.  Since $W(t)$, is a function of  $\wmax(t)$ and $s(t)$, it is completely determined by their DEs.
The result is the following model\footnote[2]{Note that $W(t)$ must be either derived explicitly, for example as in (\ref{eq:rct}) for TCP Reno or given in the definition of the controller, as in (\ref{eq:cubiccwnd}) for CUBIC.}:
\begin{align}
\label{eq:system}
\begin{split}
\frac{d\wmax(t)}{dt} &= -(\wmax(t)-W(t))\frac{W(t-\tau)}{\tau}p(t-\tau)\\
\frac{ds(t)}{dt} &= 1-s(t)\frac{W(t-\tau)}{\tau}p(t-\tau)
\end{split}
\end{align}
Above, $p(t-\tau)$ is a loss probability function. The expression $W(t-\tau)p(t-\tau)/\tau$ describes the rate of packet loss, which is delayed by $\tau$ because loss occurs at the congestion point, not at the source. The first DE describes the behavior of $\wmax$, which takes the value of $W(t)$ right before a loss. At the time of loss, if $\wmax(t)>W(t)$, then $\wmax$ decreases by the amount $\wmax(t)-W(t)$; otherwise, it increases by the same amount. The second DE describes the evolution of the time since last loss $s(t)$, which grows by one unit and is reset to zero upon loss.
\if{false}
The linearized system looks as follows:
\begin{align}
\label{sys:linearized}
\begin{split}
&\begin{bmatrix}
\delta \wmaxdot \\[0.3em]
\delta \dot{s}
\end{bmatrix} = 
\frac{1}{\shat}
A_0
\begin{bmatrix}
\delta\wmax\\[0.3em]
\delta s
\end{bmatrix}+
A_1
\begin{bmatrix}
\delta \wtau\\[0.3em]
\delta \ptau
\end{bmatrix}, \text{ where}\\
&A_0 = \begin{bmatrix}
\frac{\delta W}{\partial\wmax}\Bigr|_{\vect{X}=\vect{\hat{X}}}-1 & \frac{\delta W}{\partial s}\Bigr|_{\vect{X}=\vect{\hat{X}}}\\[0.8em]
0 & -1
\end{bmatrix},\ \ A_1=\begin{bmatrix}
0 & 0\\[0.3em]
-\frac{\shat}{\tau} & 0
\end{bmatrix},
\end{split}
\end{align}
$\vect{X}=[\wmax\ s]^T$, $\vect{\hat{X}}$ is the fixed point value of $\vect{X}$, and for a variable $v$ with fixed point value $\hat{v}$, $\delta v=v(t)-\hat{v}$.
\fi
This system can be adapted to a rate-based scheme in terms of maximum rate and time since last rate decrease, simply by dividing each DE by $\tau$. Since we will be describing applications of this model to TCP Reno and CUBIC, which are both \cwnd-based, 
we use (\ref{eq:system}) in the interest of the paper. 

Figure \ref{fig:renofig} illustrates $\wmax(t)$, $s(t)$, and $W(t)$ for TCP Reno.
To adapt model (\ref{eq:system}) to TCP Reno, we define Reno's \cwnd as a function of $\wmax(t)$ and $s(t)$. At the time of loss, $W(t)=\wmax(t)$ is halved. This becomes the initial value of $W(t)$ in the new congestion epoch. $W(t)$ then increases by one segment for every round-trip time, so the total increase is $s(t)/\tau$ after $s(t)$ time has elapsed since the last loss. Hence,
\begin{align}
W(t) &= \frac{\wmax(t)}{2}+\frac{s(t)}{\tau}.
\label{eq:rct}
\end{align}
Then the fluid model for Reno is (\ref{eq:system}) combined with (\ref{eq:rct}).

The loss probability function can be customized according to the specific characteristics of a given system, such as queue size and AQM policy. For simplicity, we use the following function in all subsequent models:
\begin{align}
p(t) &= \max{\left(1-\frac{C\tau}{W(t)},0\right)}.
\label{eq:p}
\end{align}
This function is presented in \cite{srikant2012mathematics}
as an approximation of the M/M/1/B drop probability when the buffer size $B \to \infty$.

Note that model (\ref{eq:system}) does not specify $W(t)$, and therein lies the versatility of this scheme. For a given \cwnd-based transport algorithm, the modeler need only to substitute a function describing the evolution of \cwnd over time, as we did for Reno. We demonstrate this technique again with CUBIC in Section \ref{cubic}.
This property of the model  is useful both for analyzing existing algorithms and examining the stability of new ones. 
\if{false}
In this section, we present two models for TCP Reno: the first, Model A, is the new scheme, which is the focus of this work. The second, Model B, is a well-known scheme originally presented in \cite{Misra:2000:FAN:347057.347421}. 
First, we show that the two models are mathematically equivalent. This fact serves as a proof of concept for Model A and implies that the new model can be applied to other data transport algorithms.
\subsubsection*{Model A}
\begin{align*}
\rmax(t) &= \frac{\wmax(t)}{\tau}\\
\rc(t) &= \frac{W_{\mathrm{reno}}(s(t))}{\tau}= \frac{1}{\tau}\left(\frac{s(t)}{\tau}+\frac{\wmax(t)}{2}\right)\\
\frac{d\rmax}{dt} &= -(\rmax(t)-\rc(t))\rc(t-\tau)p(t-\tau)\\
\frac{ds}{dt} &= 1-s(t)\rc(t-\tau)p(t-\tau)\\
p(t) &= \max{\left(1-\frac{C\tau}{W(t)},0\right)}
\end{align*}
\fi
\subsection{Model Equivalence for TCP Reno}
\label{reno}
\if{false}
To adapt model (\ref{eq:system}) to TCP Reno, we define Reno's \cwnd as a function of $\wmax(t)$ and $s(t)$. At the time of loss, $W(t)=\wmax(t)$ is halved. This becomes the initial value of $W(t)$ in the new congestion epoch. $W(t)$ then increases by one segment for every round-trip time, so the total increase is $s(t)/\tau$ after $s(t)$ time has elapsed since the last loss.
\begin{align}
W(t) &= \frac{\wmax(t)}{2}+\frac{s(t)}{\tau}
\label{eq:rct}
\end{align}
\fi
Consider the well-established model for TCP Reno's \cwnd from \cite{Misra:2000:FAN:347057.347421} (equation (4) to be precise):
\begin{align}
\frac{dW(t)}{dt} &= \frac{1}{\tau}-\frac{W(t)}{2}\frac{W(t-\tau)}{\tau}p(t-\tau).
\label{eq:oldmodel}
\end{align}
We assume $\tau$ to be constant for simplicity, even though the round-trip time in \cite{Misra:2000:FAN:347057.347421} varies in time as a function of both the propagation and queueing delays.

We now show that the two models (\emph{i.e.} the model represented by (\ref{eq:system}), (\ref{eq:rct}) and the model represented by (\ref{eq:oldmodel})) are mathematically equivalent. 
Differentiating (\ref{eq:rct}) with respect to $t$, we have:
\begin{align}
\dot{W} &=\frac{\wmaxdot}{2}+ \frac{\dot{s}}{\tau}.
\label{eq:rcdot}
\end{align}
Substituting (\ref{eq:system}) into (\ref{eq:rcdot}) yields
\begin{align}
\dot{W} &= \frac{1}{\tau}\left(1-s\frac{\wtau}{\tau}\ptau\right)+\frac{1}{2}\left((W-\wmax)\frac{\wtau}{\tau}\ptau\right).
\label{eq:dotw}
\end{align}

From (\ref{eq:rct}), we know that $\wmax = 2(W-s/\tau)$. Substituting this expression for $\wmax$ into (\ref{eq:dotw}) and simplifying yields
\begin{align*}
\dot{W} &= \frac{1}{\tau}\left(1-s\frac{\wtau}{\tau}\ptau\right)+\frac{1}{2}\left(\left(W-2\left(W-\frac{s}{\tau}\right)\right)\frac{\wtau}{\tau}\ptau\right)\\
&= \frac{1}{\tau}-s\frac{\wtau}{\tau^2}\ptau-\frac{W}{2}\frac{\wtau}{\tau}\ptau+s\frac{\wtau}{\tau^2}\ptau\\
&= \frac{1}{\tau}-\frac{W}{2}\frac{\wtau}{\tau}\ptau.
\end{align*}
\if{false}
We have shown that
\begin{align*}
\frac{d\rc}{dt} &= \frac{1}{\tau^2}-\frac{\rc(t)}{2}\rc(t-\tau)p(t-\tau)
\end{align*}
We can obtain an equation for the \cwnd by multiplying both sides by $\tau$:
\begin{align*}
\frac{dW}{dt} &= \frac{1}{\tau}-\frac{W(t)}{2}\frac{W(t-\tau)}{\tau}p(t-\tau) = \frac{1}{\tau}\left(1-\frac{W(t)}{2}W(t-\tau)p(t-\tau)\right)
\end{align*}
\fi
The last line corresponds to equation (\ref{eq:oldmodel}) and completes our proof of the equivalence of the models. When used with Reno, model (\ref{eq:system}) can be linearized and used to derive a transfer function. The latter can be analyzed to yield system parameter-dependent conditions for Reno's stability. This analysis is similar to the one presented in \cite{srikant2012mathematics}.
\section{Analysis of TCP CUBIC}
In this section, we perform a local stability analysis of TCP CUBIC. To do so, we first create a fluid model for this congestion control algorithm using the framework introduced in the previous section. Then, we show that the system has a unique fixed point and prove the existence and uniqueness of a solution. Next, we show that the linearization method yields inconclusive results when applied to the model, and are thus motivated to use
Lyapunov's direct method to prove the stability of the system. First, we introduce a Lyapunov function candidate and
since the system is time-delayed, use Razumikhin's Theorem to show that the candidate is suitable and that stability holds in a neighborhood of the fixed point of the system. A consequence of the failed linearization is that we will not prove exponential stability for CUBIC, but we can still show asymptotic and Lyapunov stability.
Finally, we derive convergence results on the system's solution. 
\label{cubic}
\subsection{TCP CUBIC Fluid Model}
\label{cubicmodel}
\begin{figure}[h!]
\centering
\includegraphics[width=0.5\textwidth]{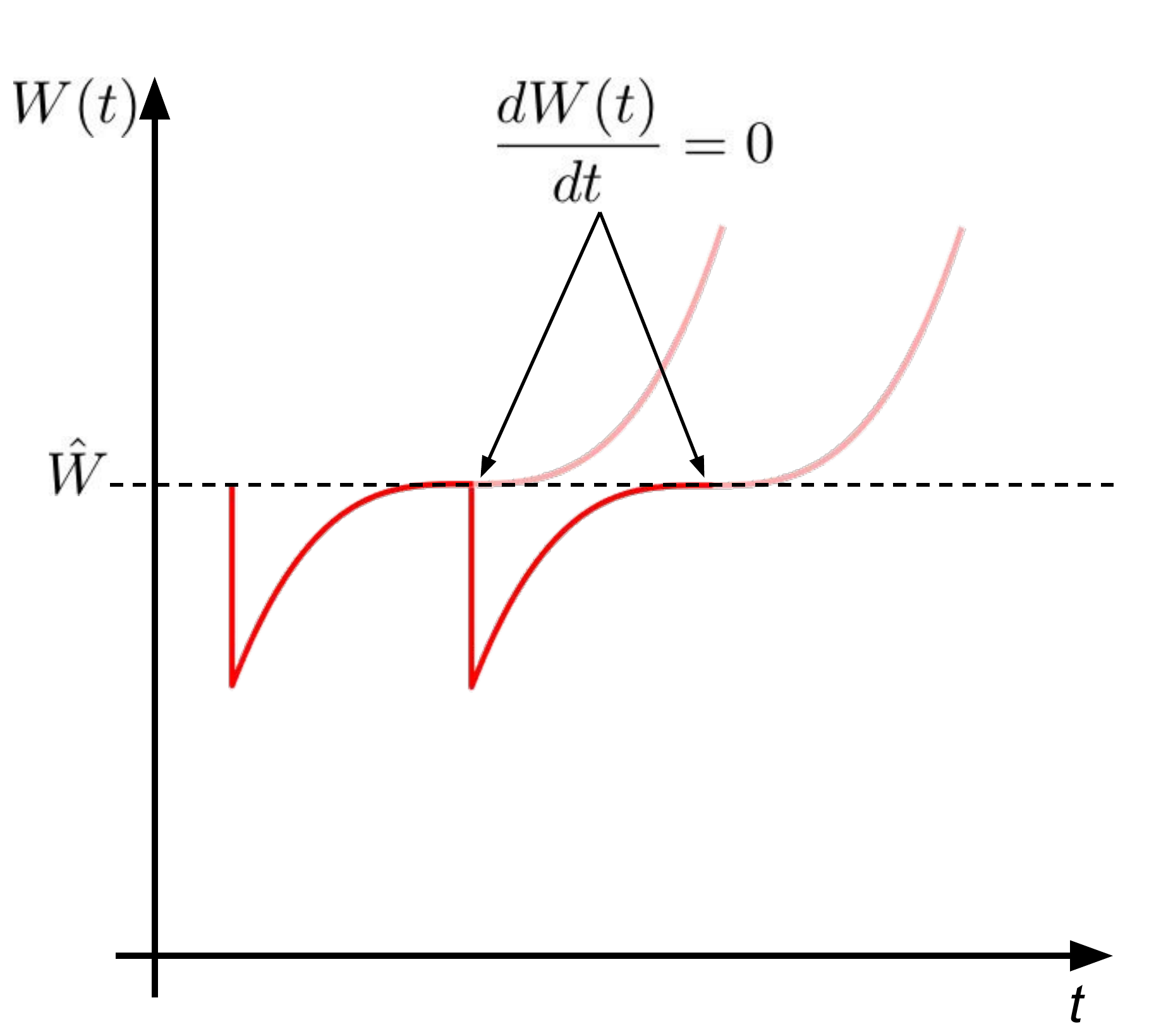}
\caption{CUBIC's saddle point causes $dW(t)/dt$ to evaluate to zero at the fixed point of the system.}
\label{fig:cubicsaddle}
\end{figure}

TCP CUBIC's congestion window function is defined in terms of the time since last loss $s(t)$ and maximum value of \cwnd immediately before the last loss $\wmax(t)$ \cite{ha2008cubic}:
\begin{align}
W(t) &= c\left(s(t)-\sqrt[3]{\frac{\wmax(t)b}{c}}\right)^3+\wmax(t)
\label{eq:cubiccwnd}
\end{align}
where $b$ is a multiplicative decrease factor and $c$ is a scaling factor. Figure \ref{fig:cubicsaddle} illustrates the evolution of CUBIC's \cwnd over time. The opaque red curves represent behavior in steady state: the window is concave until a loss occurs at CUBIC's fixed-point value of \cwnd, $\what$. The light red curves describe \cwnd behavior if a loss does not occur: the window becomes convex, also known as CUBIC's probing phase.
The fluid model for CUBIC is then simply (\ref{eq:system}) coupled with (\ref{eq:cubiccwnd}), with (\ref{eq:p}) as the loss probability function. 
Prior to the development of (\ref{eq:system}), we attempted to develop a fluid model by first computing the equilibrium point for CUBIC, but this exercise gave a value of $s$ at (\ref{eq:cubiccwnd})'s saddle point and consequently, a confounding linearization of $dW/dt=0$. Further attempts at deriving $dW/dt$, taking into account the time-dependencies $s(t)$ and $\wmax(t)$, resulted in
a highly complex DE involving both $\wmax(t)$, $s(t)$, and their derivatives. Even obtaining the fixed points of this DE would be highly cumbersome, compared to obtaining the fixed point of (\ref{eq:system}).
\subsection{Fixed Point Analysis}
Let $\wmaxhat$, $\hat{s}$, $\hat{W}$, and $\hat{p}$ represent the fixed point  values of $\wmax(t)$, $s(t)$, $W(t)$, and $p(t)$, respectively. Using the fact that in steady state, $W(t)=W(t-\tau)=\what$ and $p(t)=p(t-\tau)=\phat$, system (\ref{eq:system}) becomes
\begin{align}
-(\wmaxhat-\what)\frac{\what}{\tau}\hat{p} &= 0,\label{eq:dwmaxfixed}\\
1-\hat{s}\frac{\what}{\tau}\phat &= 0.\label{eq:dsfixed}
\end{align}
From (\ref{eq:dsfixed}), we see that
\begin{align}
\hat{s}=\frac{\tau}{\what\phat}\label{eq:shat}.
\end{align}
It is clear that $\what$ and $\phat$ do not equal zero in steady state. Using this information, along with (\ref{eq:dwmaxfixed}), we conclude that $\wmaxhat=\what$.
In steady state, (\ref{eq:cubiccwnd}) becomes
\begin{align*}
\hat{W} &= c\left(\hat{s}-\sqrt[3]{\frac{\what b}{c}}\right)^3+\what.
\end{align*}
This equation yields 
\begin{align}
\hat{s}=\sqrt[3]{\frac{\what b}{c}}\label{eq:shat2}.
\end{align}
Combining (\ref{eq:shat}) and (\ref{eq:shat2}), we have
\begin{align*}
\frac{\tau}{\what\phat} &= \sqrt[3]{\frac{\what b}{c}}
\end{align*}
where $\phat=1-C\tau/\what$ (since $\phat>0$ in steady state). Substitution yields
\begin{align*}
\hat{W}(\hat{W}-C\tau)^3 &= \frac{\tau^3c}{b},
\end{align*}
which can be solved for $\what$ as a function of solely the system parameters $c$, $b$, $C$, and $\tau$. This value can be used either with (\ref{eq:shat}) or (\ref{eq:shat2}) to obtain a value for $\shat$ solely as a function of the system parameters. This concludes the fixed point analysis. An interesting comparison is $\what$ as a function of $\phat$ for Reno and CUBIC. Model (\ref{eq:oldmodel}) yields
\begin{align*}
\what_{Reno} &= \sqrt{\frac{2}{\phat}} \text{,  while  } \what_{CUBIC} = \sqrt[4]{\frac{\tau^3c}{\phat^3b}}.
\end{align*}
In other words, whereas throughput under Reno depends on loss probability as $\mathcal{O}(\phat^{-1/2})$, CUBIC exhibits a $\phat^{-3/4}$ dependence.

\subsection{Change of Variables}
To simplify stability analysis, we perform a change of variables so that the fixed point of the system is located at the origin.
To accomplish this, define $\vect{x}$ as follows:
\begin{align*}
\vect{x}(t) &=
 \begin{bmatrix}
x_1(t)\\[0.3em]
x_2(t)
\end{bmatrix} =
\begin{bmatrix}
\wmax(t)-\wmaxhat\\[0.3em]
s(t)-\hat{s}
\end{bmatrix}=
\begin{bmatrix}
\wmax(t)-\what\\[0.3em]
s(t)-\hat{s}
\end{bmatrix}
\end{align*}
where the last equality follows because $\wmaxhat=\what$.
Also, define $\Psi(t)$ and $\tilde{p}(t)$ as follows:
\begin{align*}
\Psi(t) &= c\left(x_2(t)+\hat{s}-\sqrt[3]{\frac{b(x_1(t)+\what)}{c}}\right)^3+x_1(t)+\what,\\
\tilde{p}(t) &= \max{\left(1-\frac{C\tau}{\Psi(t)},0\right)}.
\end{align*}
Then the new system is:
\begin{align}
\label{sys:model}
\begin{split}
\xodot &= \left(\Psi-x_1-\what\right)\frac{\psitau}{\tau}\ptildetau,\\
\xtdot &= 1-(x_2+\hat{s})\frac{\psitau}{\tau}\ptildetau.
\end{split}
\end{align}
Note that $\psitau$ and $\ptildetau$ are functions of ${x_1(t-\tau)\equiv \xotau}$ and ${x_2(t-\tau)\equiv \xttau}$. 
It is easy to verify that ${\vect{x^*}=[x_1\ x_2\ \xotau\ \xttau]^T=\vect{0}}$ is a fixed point of the new system.
\begin{claim} $\vect{x^*}=\vect{0}$ is a fixed point of the new system.\\
\noindent\textbf{Proof:} At $\vect{x^*}$, we have:
\begin{align}
\hat{\Psi} &= c\left(\hat{s}-\sqrt[3]{\frac{b\what}{c}}\right)^3+\what \nonumber\\
\xodot &= \left(\Psi-\what\right)\frac{\psitau}{\tau}\ptildetau \label{eq:1}\\
\xtdot &= 1-\hat{s}\frac{\psitau}{\tau}\ptildetau \label{eq:2}
\end{align}
From the fixed point analysis of the original system, recall that $\hat{s}=\sqrt[3]{b\what/c}$, which yields $\hat{\Psi}=\what$. Plugging this into equation (\ref{eq:1}), we get $\xodot=0$. Similarly, plugging in $\hat{\Psi}=\what$ into (\ref{eq:2}), we have:
\begin{align*}
\xtdot &= 1-\hat{s}\frac{\what}{\tau}\hat{p}
\end{align*}
From the fixed point analysis of the original system, recall that $\hat{s}=\tau/(\what\hat{p})$. Therefore, $\dot{x_2}=0$ as well, and the proof is complete. \qed
\end{claim}
\if{false}
\emph{Claim:} $\vect{x^*}=\vect{0}$ is a fixed point of the new system.\\
\emph{Proof:} At $\vect{x^*}$, we have:
\begin{align}
\hat{\Psi} &= c\left(\hat{s}-\sqrt[3]{\frac{b\what}{c}}\right)^3+\what \nonumber\\
\xodot &= \left(\Psi-\what\right)\frac{\psitau}{\tau}\ptildetau \label{eq:1}\\
\xtdot &= 1-\hat{s}\frac{\psitau}{\tau}\ptildetau \label{eq:2}
\end{align}
From the fixed point analysis of the original system, recall that $\hat{s}=\sqrt[3]{b\what/c}$, which yields $\hat{\Psi}=\what$. Plugging this into equation (\ref{eq:1}), we get $\xodot=0$. Similarly, plugging in $\hat{\Psi}=\what$ into (\ref{eq:2}), we have:
\begin{align*}
\xtdot &= 1-\hat{s}\frac{\what}{\tau}\hat{p}
\end{align*}
From the fixed point analysis of the original system, recall that $\hat{s}=\tau/(\what\hat{p})$. Therefore, $\dot{x_2}=0$ as well, and the proof is complete.
\fi
We can analyze the stability of the system (\ref{sys:model}) at the origin, which is equivalent to analyzing the stability of the original system (\ref{eq:system}) at the equilibrium values $\wmaxhat$ and $\hat{s}$. 
The CUBIC representation in (\ref{sys:model}) forms the basis for our subsequent analyses.
\subsection{Existence and Uniqueness of Solution}
\label{exuniq}
We state the existence and uniqueness theorem at it appears in \cite{gu2003stability}:
\begin{theorem}[Theorem 1.2 from \emph{Stability of Time-Delay Systems}]
Suppose that $\Omega$ is an open set in $\mathbb{R}\times \mathcal{C}$ (where $\mathcal{C}$ is the set of $\mathbb{R}^n$-valued continuous functions on $[-\tau, 0]$), 
$f:\Omega \to \mathbb{R}^n$ is continuous, and $f(t,\phi)$ is Lipschitzian in $\phi$ in each compact set in $\Omega$, that is, for each given compact set $\Omega_0 \subset \Omega$, there exists a constant $L$ such that
\begin{align*}
||f(t,\phi_1)-f(t,\phi_2)|| \leq L||\phi_1-\phi_2||
\end{align*}
for any $(t,\phi_1)\in\Omega_0$ and $(t,\phi_2)\in\Omega_0$. If $(t_0,\phi)\in\Omega$, then there exists a unique solution of $\dot{x}(t)=f(t,x_t)$ through $(t_0,\phi)$.
\end{theorem}
To prove existence and uniqueness for our system, it is sufficient to show that $\xodot$ and $\xtdot$ are continuously differentiable functions in some neighborhood of the fixed point. We assume that this neighborhood is small enough so that $\ptildetau>0$. Then the system becomes:
\begin{align*}
\xodot &= \left(\Psi-x_1-\what\right)\frac{(\psitau-C\tau)}{\tau},\\
\xtdot &= 1-(x_2+\hat{s})\frac{(\psitau-C\tau)}{\tau}.
\end{align*}
\begin{align*}
\text{Let } F=\frac{b(x_1+\what)}{c} \text{ and } 
\Phi=x_2+\shat   - \sqrt[3]{F}.
\end{align*}
Following are the partial derivatives of $\xodot$:
\begin{align*}
\frac{\partial \xodot}{\partial x_1} &= -\frac{b}{\tau}\Phi^2F^{-2/3}(\psitau-C\tau),\\
\frac{\partial \xodot}{\partial x_2} &= \frac{3c}{\tau}\Phi^2(\psitau-C\tau),\\
\frac{\partial \xodot}{\partial \xotau} &=\frac{(\Psi-x_1-\what)}{\tau}\left(-b\Phi_{\tau}^2F_{\tau}^{-2/3}+1\right),\\
\frac{\partial \xodot}{\partial \xttau} &=\frac{3c}{\tau}(\Psi-x_1-\what)\Phi_{\tau}^2.
\end{align*}
These partials provide the first restriction to the region where stability is being analyzed. Specifically, 
the term $F^{-2/3}$ indicates that 
$x_1$ and $\xotau$ should be restricted to 
an interval $[-\rho\what,\rho\what]$, $0<\rho<1$. 
Next, we look at the partial derivatives of $\xtdot$:
\begin{align*}
\frac{\partial \xtdot}{\partial x_1} &= 0,\ \frac{\partial \xtdot}{\partial x_2} =-\frac{(\psitau-C\tau)}{\tau},\\
\frac{\partial \xtdot}{\partial \xotau} &= -\frac{(x_2+\shat)}{\tau}\left(-b\Phi_{\tau}^2F_{\tau}^{-2/3}+1\right),\\
\frac{\partial \xtdot}{\partial \xttau} &= -\frac{3c}{\tau}(x_2+\shat)\Phi_{\tau}^2.
\end{align*}
Under the restriction stated above, these partials are also continuous, and hence, we have local Lipschitz continuity -- the requirement for existence and uniqueness.
\subsection{Stability Analysis}
\noindent In general, the linearization of (\ref{sys:model}) and (\ref{eq:p}) about $\vect{x}=\vect{x^*}=\vect{0}$ is
\begin{align*}
\begin{bmatrix}\dot{x}_1\\[0.3em]
\dot{x}_2\end{bmatrix}
 &= \frac{1}{\shat}A_0\begin{bmatrix}x_1\\[0.3em]
x_2\end{bmatrix}-\frac{\shat}{\tau}A_1\begin{bmatrix}\xotau\\[0.3em]
\xttau\end{bmatrix},\text{ where } \\
A_0&=\begin{bmatrix}\frac{\partial \Psi}{\partial x_1}-1 & \frac{\partial \Psi}{\partial x_2}\\[0.3em]
0 & -1
\end{bmatrix}\Bigg |_{\vect{x}=\vect{x^*}}
\text{ and } A_1=\begin{bmatrix}
0 & 0\\[0.3em]
\frac{\partial \psitau}{\partial \xotau} & \frac{\partial \psitau}{\partial \xttau}
\end{bmatrix}\Bigg |_{\vect{x}=\vect{x^*}}.
\end{align*}
\if{false}
Linearizing system (\ref{sys:model}) about $\vect{x}=\vect{x^*}=\vect{0}$, we obtain
\begin{align*}
\begin{bmatrix}\dot{x}_1\\[0.3em]
\dot{x}_2\end{bmatrix}
 &= A_0\begin{bmatrix}x_1\\[0.3em]
x_2\end{bmatrix}+A_1\begin{bmatrix}\xotau\\[0.3em]
\xttau\end{bmatrix},\text{ where } \\
A_0&=\begin{bmatrix}0 & 0\\[0.3em]
0 & -\frac{1}{\shat}
\end{bmatrix}\text{ and } A_1=\begin{bmatrix}
0 & 0\\[0.3em]
-\frac{\shat}{\tau} & 0
\end{bmatrix}.
\end{align*}
\fi
For CUBIC, $\partial \Psi/\partial x_1|_{\vect{x}=\vect{x^*}}=1$ and $\partial \Psi/\partial x_2|_{\vect{x}=\vect{x^*}}=0$, so that
$\xodot=0$. This means that linearization has failed; \emph{i.e.}, stability of the linearized system cannot be generalized to local stability of the nonlinear system.
The key cause of this problem is the fact that the fixed point value of $x_2$, $0$, is the saddle point of the function $\Psi$ (or equivalently, $\shat$ is the saddle point of $W(t)$). 
Figure \ref{fig:cubicsaddle} illustrates this phenomenon.
This causes all first-order partial derivatives of $\xodot$ to evaluate to zero at $\vect{x^*}=\vect{0}$.
Hence, in order to incorporate a local contribution from $\xodot$ in the analysis, it is necessary to expand $\xodot$ further. Specifically, a third-order Taylor Series expansion is necessary, since all second-order terms also evaluate to zero at the origin.

The expanded system looks as follows:
\begin{align}
\label{sys:exp}
\begin{split}
\xodot &= -\alpha x_1^3+\beta x_1^2x_2-\gamma x_1x_2^2+\delta x_2^3+h_1\\
\xtdot &= -\frac{1}{\hat{s}}x_2-\frac{\hat{s}}{\tau}\xotau+h_2 
\end{split}
\end{align}
\vspace{-1em}
\begin{align*}
\text{where } \alpha=\frac{b^3}{27c^2\hat{s}^7},\ \beta=\frac{b^2}{3c\hat{s}^5},\ \gamma=\frac{b}{\hat{s}^3},\text{ and } \delta=\frac{c}{\hat{s}}
\end{align*}
and $h_1$ and $h_2$ are higher-order terms of $\xodot$ and $\xtdot$, respectively.
To analyze the stability of (\ref{sys:exp}), we use the Lyapunov-Razumikhin Theorem, the statement of which is given below as it appears in \cite{gu2003stability}. For the purpose of this theorem, 
we introduce some notation. Let $\mathcal{C}=\mathcal{C}([-\tau,0],\mathbb{R}^n)$ be the set of continuous functions mapping the interval $[-\tau,0]$ to $\mathbb{R}^n$, where $\tau$ is the maximum delay of a system.
For any $A>0$ and any continuous function of time $\psi \in \mathcal{C}([t_0-\tau,t_0+A],\mathbb{R}^n)$, and $t_0\leq t\leq t_0+A$, let $\psi_t \in \mathcal{C}$ be a segment of the function $\psi$ defined as $\psi_t(\theta)=\psi(t+\theta)$, $-\tau\leq\theta\leq 0$.
The general form of a retarded functional differential equation is
\begin{align}
\dot{x}(t)=f(t,x_t) \label{eq:rfde}
\end{align}
Below, $\mathbb{R}_+$ is the set of positive real numbers, and $\bar{\mathbb{S}}$ is the closure of the set $\mathbb{S}$.
\begin{theorem}[Lyapunov-Razumikhin Theorem]\label{thm}
Suppose $f: \mathbb{R}\times \mathcal{C}\to \mathbb{R}^n$ takes $\mathbb{R}\times$(bounded sets of $\mathcal{C}$) into bounded sets of $\mathbb{R}^n$, and $u$, $v$, $w: \bar{\mathbb{R}}_+\to\bar{\mathbb{R}}_+$ are continuous nondecreasing functions, $u(s)$ and $v(s)$ are positive for $s>0$, and $u(0)=v(0)=0$, $v$ strictly increasing. If there exists a continuously differentiable function $V:\mathbb{R}\times\mathbb{R}^n\to\mathbb{R}$ such that
\begin{align}
u(||x||) \leq V(t,x) \leq v(||x||),\ \text{ for } t\in \mathbb{R} \text{ and } x\in\mathbb{R}^n,
\label{eq:Vbounds}
\end{align}
$w(s)>0$ for $s>0$, and there exists a continuous nondecreasing function $p(s)>s$ for $s>0$ such that 
\begin{align}
\dot{V}(t,x(t)) \leq -w(||x(t)||)\label{eq:Vdotw}\\ 
\text{ if } V(t+\theta,x(t+\theta))\leq p(V(t,x(t)))
\label{eq:Vdotcondstr}
\end{align}
for $\theta \in [-\tau,0]$, then the system (\ref{eq:rfde}) is uniformly asymptotically stable.
If in addition $\lim_{s\to\infty}u(s)=\infty$, then the system (\ref{eq:rfde}) is globally uniformly asymptotically stable.
\end{theorem}
Note that in this work, we will only prove \emph{local} stability for CUBIC. Therefore, our goal is to show that we can find a function $V$ for which all conditions specified in the theorem are valid locally, \emph{i.e.}, in a sufficiently small neighborhood around the fixed point.

A popular choice of Lyapunov function is the quadratic candidate, \emph{i.e.} a function of the form
\begin{align}
Z(\vect{x}) &= \vect{x}^TP\vect{x}, \text{ where } P=\begin{bmatrix}
p_1 & p_2\\[0.3em]
p_2 & p_4
\end{bmatrix}
\label{eq:lyapquad}
\end{align}
is positive definite.
 Not surprisingly, the quadratic form $Z$, which is a sufficient form in working with linear dynamic systems, proves unsuitable.
 To understand why, consider the time derivative of (\ref{eq:lyapquad}) along solutions to (\ref{sys:exp}):
\begin{align}
\dot{Z} &= 2\xodot(p_1x_1+p_2x_2)+2\xtdot(p_2x_1+p_4x_2).
\label{eq:wdot}
\end{align}
The first term above is quartic in $x_1$ and $x_2$ (because $\xodot$ in (\ref{sys:exp}) is cubic in $x_1$, $x_2$), but the second term is quadratic in $x_1$, $x_2$, and $\xotau$. In a small neighborhood of $\vect{x^*}$, the quadratic terms dominate; \emph{i.e.}, (\ref{eq:wdot}) becomes
\begin{align*}
\dot{Z} &= 2\left(-\frac{1}{\hat{s}}x_2-\frac{\hat{s}}{\tau}\xotau\right)(p_2x_1+p_4x_2)+h.o.t.,
\end{align*}
where $h.o.t.$ denotes higher-order terms. We cannot guarantee negativity of these terms, even locally. The main problem with $\dot{Z}$ is that $p_2$ must be non-zero for $\dot{Z}$ to be negative definite, yet this is the same coefficient responsible for the cross term of $x_1$ and $x_2$ in (\ref{eq:lyapquad}), which prevents us from effectively bounding $\xotau$ using condition (\ref{eq:Vdotcondstr}) of Theorem \ref{thm}.
 However, the failure of this quadratic Lyapunov function serves as an instructive example. Namely, we would like a Lyapunov candidate to have the following two properties: (i) it must prevent $\xtdot$'s terms from dominating the Lyapunov derivative, and (ii) the cross terms of $x_1$ and $x_2$ in the Lyapunov function should be absent so that delayed terms (like $\xotau$) can be easily bounded with non-delayed versions (like $x_1$) using (\ref{eq:Vdotcondstr}). With these motivations,
consider the following Lyapunov-Razumikhin candidate:
\begin{align}
\label{eq:lyapfunc}
V(\vect{x}) &= \frac{d_1}{2}x_1^2+\frac{d_4}{4}x_2^4
\end{align}
where $d_1$ and $d_4$ are positive. In a subsequent discussion, we will specify the values of $d_1$ and $d_4$ in terms of system parameters. We will also show that $V$ can be bounded by functions $v(||\vect{x}||)=\epsilon_0||\vect{x}||_2^2$ and $u(||\vect{x}||)=\epsilon_1||\vect{x}||_2^4$, for appropriate choices of constants $\epsilon_0$ and $\epsilon_1$, and that these functions satisfy all conditions specified in Theorem \ref{thm}.
$V$ satisfies (ii), as necessary, and allows us to choose a convenient function $p(V(\vect{x}(t)))$ for (\ref{eq:Vdotcondstr}) (note: we can write $V(\vect{x}(t))$ instead of $V(t,\vect{x}(t))$ because $V$ is autonomous, \emph{i.e.} it is not explicitly a function of time). Let $p>1$ be a constant, which can be arbitrarily close to one. Then for (\ref{eq:Vdotcondstr}), we can use $p(V(\vect{x}(t)))=pV(\vect{x}(t))$:
\begin{align*}
V(\vect{x}(t-\theta))\leq pV(\vect{x}(t)), \text{ for } \theta\in[0,\tau].
\end{align*}
Since there are no cross terms of $\xotheta$ and $\xttheta$ on the left-hand side of the above inequality, bounding the absolute values of these delayed variables individually is straightforward (and instrumental in the proofs that follow).

Now, consider the Lyapunov derivative:
\begin{align*}
\dot{V} &= d_1x_1\xodot +d_4x_2^3\xtdot.
\end{align*}
Note that both of the terms above are now \emph{quartic} in either $x_1$, $x_2$, or both.  However, $\xtdot$ still contributes a term with $\xotau$, which poses a challenge in proving local stability. Indeed, at the core of the proof for $\dot{V}$'s negativity is managing the $\xotau$ term, as well as proving that $h_1$ and $h_2$ in (\ref{sys:exp}), which are higher-order in both the delayed and non-delayed variables, are also higher-order in \emph{only} the non-delayed variables $x_1$ and $x_2$.

The focus of the next discussion is the term that contains $\xotau$.
Substituting the expanded system (\ref{sys:exp}) into $\dot{V}$ and rearranging terms, we have
\begin{align*}
\dot{V} &= d_1x_1\left(-\alpha x_1^3+\beta x_1^2x_2-\gamma x_1x_2^2+\delta x_2^3\right) +d_4x_2^3\left(-\frac{1}{\hat{s}}x_2-\frac{\hat{s}}{\tau}\xotau\right)+d_1x_1h_1+d_4x_2^3h_2\\
&= d_1\left(-\alpha x_1^4+\beta x_1^3x_2-\gamma x_1^2x_2^2+\delta x_1x_2^3\right)-\frac{d_4}{\hat{s}}x_2^4-d_4\frac{\hat{s}}{\tau}x_2^3\xotau+d_1x_1h_1+d_4x_2^3h_2\\
&= d_1\left(-\alpha x_1^4+\beta x_1^3x_2-\gamma x_1^2x_2^2\right)-\frac{d_4}{\hat{s}}x_2^4+d_1\delta x_1x_2^3-d_4\frac{\hat{s}}{\tau}x_2^3\xotau+d_1x_1h_1+d_4x_2^3h_2\\
&= \vect{y}^TQ\vect{y}+d_1\delta x_1x_2^3-d_4\frac{\hat{s}}{\tau}x_2^3\xotau+d_1x_1h_1+d_4x_2^3h_2
\end{align*}
\begin{align*}
\text{where } \vect{y}=
\begin{bmatrix}
x_1^2 \\[0.3em]
x_1x_2 \\[0.3em]
x_2^2
\end{bmatrix} \text{ and }
Q =
\begin{bmatrix}
-d_1\alpha & d_1\beta/2 & 0\\[0.3em]
d_1\beta/2 & -d_1\gamma & 0\\[0.3em]
0 & 0 & -d_4/\shat
\end{bmatrix}
\end{align*}

\noindent Recall the Mean Value Theorem.
\begin{theorem}[Mean Value Theorem]
Let ${\displaystyle f:[a,b]\to \mathbb {R} }$ be a continuous function on the closed interval ${\displaystyle [a,b]}$, and differentiable on the open interval ${\displaystyle (a,b)}$, where ${\displaystyle a<b}$. Then there exists some ${\displaystyle c}$ in ${\displaystyle (a,b)}$ such that
\begin{align*}
{\displaystyle f'(c)={\frac {f(b)-f(a)}{b-a}}.}
\end{align*}
\end{theorem}
Let $I$ be the interval $[t-\tau,t]$. Then by the MVT, there exists some $\theta\in(0,\tau)$ such that
\begin{align*}
\xodot(t-\theta) &=\frac{x_1(t)-x_1(t-\tau)}{t-(t-\tau)}=\frac{x_1(t)-x_1(t-\tau)}{\tau},\\
x_1(t-\tau) &= x_1(t)-\xodot(t-\theta)\tau,\ \theta\in(0,\tau),\\
\text{or } \xotau &= x_1-\dot{x}_{1_{\theta}}\tau,
\end{align*}
where $\dot{x}_{1_{\theta}}=\xodot(t-\theta)$.
\noindent We would like to combine the
terms $d_1\delta x_1x_2^3$ and $-d_4\frac{\hat{s}}{\tau}x_2^3\xotau$ in $\dot{V}$
using the MVT. To do so, let $d_1=1/\delta=\shat/c$ and $d_4=\tau/\shat$.
\begin{align*}
\dot{V} &=\vect{y}^TQ\vect{y}+x_1x_2^3-x_2^3\xotau+d_1x_1h_1+d_4x_2^3h_2\\
&=\vect{y}^TQ\vect{y}+x_2^3(x_1-\xotau)+d_1x_1h_1+d_4x_2^3h_2\\
&=\vect{y}^TQ\vect{y}+x_2^3\dot{x}_{1_{\theta}}\tau+d_1x_1h_1+d_4x_2^3h_2
\end{align*}
Note that the last three terms above all have dependencies on $\xotau$ and $\xttau$. Our goal is to show that these terms are of higher order than $\vect{y}^TQ\vect{y}$ in variables $x_1$ and $x_2$ alone.
Consider $\dot{x}_{1_{\theta}}$:
\begin{align*}
\dot{x}_{1_{\theta}} &= \Phi_{\theta}^3\frac{\Psi_{\theta+\tau}}{\tau}\tilde{p}_{\theta+\tau} 
\end{align*}
We would like to find an upper bound for $|\dot{x}_{1_{\theta}}|$ only in terms of $x_{1_{\theta}}$ and $x_{2_{\theta}}$.
Since $\theta\in(0,\tau)$, we can expand $\Phi_{\theta}^3$ about $\vect{v}=[x_{1_{\theta}}\ x_{2_{\theta}}]=[0\ 0]$
(in other words, the fixed point implicitly includes all $x_1(t-\xi)$, $x_2(t-\xi)$, $0\leq\xi\leq\tau$, not just $x_1(t)$, $x_2(t)$, $x_1(t-\tau)$, and $x_2(t-\tau)$).
Note also that by performing this expansion, we are applying a two-variable Taylor series expansion to a four-variable function. 
Specifically, we can use a second-order expansion and bound $|\dot{x}_{1_{\theta}}|$ using only the remainder, which consists of third-order partial derivatives.

In the expressions that follow, we use $F$ and $\Phi$ as defined in subsection \ref{exuniq}. Also, let $\Gamma = \Psi_{\theta+\tau}\tilde{p}_{\theta+\tau}/\tau$. The zero-, first-, and second-order terms in the expansion of 
$\dot{x}_{1_{\theta}}$ are zero when evaluated at $\vect{v}=\vect{0}$.
The third-order partial derivatives are:
\begin{align*}
\frac{\partial \dot{x}_{1_{\theta}}(\vect{v}=\vect{0})}{\partial \xotheta^3} &= \frac{-2b^3}{9c^2}\Gamma\left[
F_{\theta}^{-2}+6\Phi_{\theta}F_{\theta}^{-7/3}+5\Phi_{\theta}^2F_{\theta}^{-8/3}
\right],\\
\frac{\partial \dot{x}_{1_{\theta}}(\vect{v}=\vect{0})}{\partial \xotheta^2\xttheta} &= \frac{2b^2}{3c}\Gamma\left[
F_{\theta}^{-4/3}+2\Phi_{\theta}F_{\theta}^{-5/3}
\right],\\
\frac{\partial \dot{x}_{1_{\theta}}(\vect{v}=\vect{0})}{\partial\xotheta\xttheta^2} &= -2b\Gamma F_{\theta}^{-2/3},\text{ and }
\frac{\partial \dot{x}_{1_{\theta}}(\vect{v}=\vect{0})}{\partial \xttheta^3} = 6c\Gamma.
\end{align*}

\noindent We will bound the absolute values of these partial derivatives 
and use the following proposition \cite{folland2005higher}.
 \begin{proposition}[]If  a function $f$ is of class $C^{k+1}$ on an open convex set $S$ and $|\partial^{\alpha}f(\vect{x})|\leq M$ for $\vect{x}\in S$ and $|\alpha|=k+1$, then the absolute value of the remainder $R_{\vect{a},k}(\vect{h})$ of the $k$th-order Taylor series expansion of $f$ about the point $\vect{a}$ can be bounded as follows:
 \begin{align*}
 |R_{\vect{a},k}(\vect{h})| &\leq \frac{M}{(k+1)!}\|\vect{h}\|^{k+1}, \text{ where}\\
 \|\vect{h}\| &=|h_1|+|h_2|+\cdots+|h_n|
 \end{align*}
 \end{proposition}
 Above, $\partial f^{\alpha}$ is the generic $(k+1)$th-order partial derivative of $f$, and $|\alpha|=\alpha_1+\alpha_2+\cdots+\alpha_n$. 
 In our case, $\vect{a}=\vect{0}$, and 
 $\vect{h}=[\xotheta\ \xttheta]$.

\if{false}
Assuming that $x_1(t)$ and $x_2(t)$ are constrained in a region $(r_{11},r_{12})$ and $(r_{21},r_{22})$, respectively, $\forall t$, we can bound $\frac{\Psi_{\theta+\tau}}{\tau}\tilde{p}_{\theta+\tau}$ first. Hence, there exists a positive constant, say $M_1$, such that
\begin{align*}
\frac{\Psi_{\theta+\tau}}{\tau}\tilde{p}_{\theta+\tau} &\leq M_1\\
\rightarrow |\dot{x}_{1_{\theta}}| &\leq |\Phi_{\theta}|M_1
\end{align*}
\fi

In three of these partial derivatives, there are terms of the form $\left(\frac{b(x_{1_{\theta}}+\what)}{c}\right)^{-l}$, where $l$ is a positive rational number. Hence, we must bound $\xotheta$ in a region $[-\rho\what,\rho\what]$, where $0< \rho <1$. Assuming that $\xotheta$, $\xttheta$, $x_{1_{\theta+\tau}}$, and $x_{2_{\theta+\tau}}$ are constrained in an appropriately-chosen local region $[-r,r]$ around $\vect{0}$, there exists a constant $M$ such that $|\partial^3 f|\leq M$. Then, using the proposition,
\begin{align}
|\dot{x}_{1_{\theta}}| &\leq \frac{M}{3!}(|x_{1_{\theta}}|+|x_{2_{\theta}}|)^3
\label{eq:x1thetadot}
\end{align}

\noindent By Razumikhin's Theorem, we require that $\dot{V}(\vect{x})\leq -w(\|\vect{x}\|)$ whenever $pV(\vect{x}(t))\geq V(\vect{x}(t-\theta))$, $\theta\in(0,\tau)$, for an $\epsilon>0$ and some constant $p>1$. When $pV(\vect{x}(t))\geq V(\vect{x}(t-\theta))$,
\begin{align*}
&p\left(\frac{d_1}{2}x_1^2+\frac{d_4}{4}x_2^4\right) \geq \frac{d_1}{2}x_{1_{\theta}}^2+\frac{d_4}{4}x_{2_{\theta}}^4\\
&p\left(d_1x_1^2+\frac{d_4}{2}x_2^4\right) \geq d_1x_{1_{\theta}}^2\\
&|x_{1_{\theta}}| \leq \sqrt{\frac{p}{d_1}}\sqrt{\left(d_1x_1^2+\frac{d_4}{2}x_2^4\right)} \leq \sqrt{\frac{p}{d_1}}\left(\sqrt{d_1x_1^2}+\sqrt{\frac{d_4}{2}x_2^4}\right)\leq
\sqrt{\frac{p}{d_1}}\left(\sqrt{d_1}|x_1|+\sqrt{\frac{d_4}{2}}x_2^2\right)
\end{align*}

Similarly,
\begin{align*}
&p\left(d_1x_1^2+\frac{d_4}{2}x_2^4\right) \geq \frac{d_4}{2}x_{2_{\theta}}^4\\
&\xttheta^4 \leq \frac{2p}{d_4}\left(d_1x_1^2+\frac{d_4}{2}x_2^4\right)\\
&|x_{2_{\theta}}| \leq \sqrt[4]{\frac{2p}{d_4}}\sqrt[4]{\left(d_1x_1^2+\frac{d_4}{2}x_2^4\right)}\leq
\sqrt[4]{\frac{2p}{d_4}}\left(\sqrt[4]{d_1x_1^2}+\sqrt[4]{\frac{d_4}{2}x_2^4}\right)\leq
\sqrt[4]{\frac{2p}{d_4}}\left(\sqrt[4]{d_1}|x_1|^{1/2}+\sqrt[4]{\frac{d_4}{2}}|x_2|\right)
\end{align*}
Substituting these results into (\ref{eq:x1thetadot}), we have:
\begin{align*}
|\dot{x}_{1_{\theta}}| &\leq \frac{M}{6}\left(\sqrt{\frac{p}{d_1}}\left(\sqrt{d_1}|x_1|+\sqrt{\frac{d_4}{2}}x_2^2\right)+\sqrt[4]{\frac{2p}{d_4}}\left(\sqrt[4]{d_1}|x_1|^{1/2}+\sqrt[4]{\frac{d_4}{2}}|x_2|\right)\right)^3\\
&= \frac{M}{6}\left(\sqrt{p}|x_1|+\sqrt{\frac{pd_4}{2d_1}}x_2^2+\sqrt[4]{\frac{2pd_1}{d_4}}|x_1|^{1/2}+\sqrt[4]{p}|x_2|\right)^3
\end{align*}
The lowest-order term in the equation above is $(|x_1|^{1/2})^3=|x_1|^{3/2}$.
Hence, we see that the term $|x_2^3\dot{x}_{1_{\theta}}\tau|$ can be bounded by a function of order at least 4.5. We can use a similar procedure to bound the remainders of $\xodot$ and $\xtdot$, as we will demonstrate. 

First, consider the higher-order terms in $\xodot$. Since $\xttheta$ depends on $\sqrt{|x_1|}$, we cannot simply use a third-order expansion of $\xodot$ and bound the remainder of fourth-order partials, because a consequence of this is that the remainder will have a term $\left(|x_1|^{1/2}\right)^4=|x_1|^2$. Recall that in the Lyapunov derivative, $\xodot$ is being multiplied by $x_1$, so the resulting term will have an order of merely three. Using this logic, it is clear that we need an expansion of at least order six; this way, the lowest-order term in the remainder will be $(|x_1|^{1/2})^7=|x_1|^{7/2}$, and $x_1|x_1|^{7/2}$ is order 4.5, which is sufficient. However, it is not enough to simply do a sixth-order expansion of $\xodot$: we must also ensure that any fourth-, fifth-, and sixth-order partial derivatives in the expansion in terms of $x_1$, $x_2$, $\xotau$, and $\xttau$, are of order 3.5 or more in terms of only $x_1$ and $x_2$ (so that when multiplied by $x_1$ in the Lyapunov derivative, we have terms of order at least 4.5).
\begin{claim}{Except for the third-order terms, the sixth-order expansion of $\xodot$ in $[x_1\ x_2\ \xotau\ \xttau]$ is of combined power at least 3.5 in $[x_1\ x_2]$.}

\noindent\textbf{Proof:} Consider the un-expanded $\xodot$ and assume that $\ptildetau>0$ in the region where we are considering this function:
\begin{align*}
\dot{x}_{1} &= \left(\Psi-x_1-\what\right)\frac{\psitau}{\tau}\ptildetau\\
&= \frac{c}{\tau}\left(x_{2}+\hat{s}-\sqrt[3]{\frac{b(x_{1}+\what)}{c}}\right)^3\left(c\left(\xttau+\hat{s}-\sqrt[3]{\frac{b(\xotau+\what)}{c}}\right)^3+\xotau+\what-C\tau\right)
\end{align*}
Recall that when expanding this function about $[x_1\ x_2\ \xotau\ \xttau]=\vect{0}$, the zero-, first-, and second-order terms are zero. 
The fourth-order terms are:
\begin{align*}
\frac{b^4x_1^4}{27c^3\shat^{10}} + \frac{cx_2^3\xotau}{\tau} - \frac{2b^3x_1^3x_2}{9c^2\shat^8} + \frac{b^2x_1^2x_2^2}{3c\shat^6} - \frac{bx_1x_2^2\xotau}{\shat^2\tau} - \frac{b^3x_1^3\xotau}{27c^2\shat^6\tau} + \frac{b^2x_1^2x_2\xotau}{3c\shat^4\tau}
\end{align*}
We see that there are no terms that depend on $\xttau$ above. The terms that contain $\xotau$ are not problematic: we can take their absolute values and replace $|\xotau|$ by an expression that depends on $|x_1|$ and $x_2^2$ using Razumikhin's Theorem, to obtain an upper-bound for these terms. We conclude that these terms have power at least four in $[x_1\ x_2]$.

Next, consider the fifth-order terms in the expansion:
\begin{align*}
\frac{13b^4x_1^4x_2}{81c^3\shat^{11}} - \frac{8b^5x_1^5}{243c^4\shat^{13}} - \frac{5b^3x_1^3x_2^2}{27c^2\shat^9} + \frac{b^4x_1^4\xotau}{27c^3\shat^9\tau} - \frac{2b^3x_1^3x_2\xotau}{9c^2\shat^7\tau} + \frac{b^2x_1^2x_2^2\xotau}{3c\shat^5\tau}
\end{align*}
Again, there are no terms that depend on $\xttau$. Terms that contain $\xttau$ only begin to show up in the sixth-order partial derivatives evaluated at $\vect{0}$. However, these terms contain at most $\xttau^3$, since taking the derivative of $\xodot$ with respect to $\xttau$ four times yields zero. The rest of the variables in such a term is any cubic combination of $x_1$, $x_2$, and $\xotau$. Hence, the minimum combined power of such a term (after taking the absolute value and bounding using Razumikhin's) is $3+3(1/2)=3+1.5=4.5>4$ in $[x_1\ x_2]$.

Finally, we can use the proposition to bound the remainder. For some positive constant $M_1$,
\begin{align*}
|R_{\vect{0},6}| \leq \frac{M_1}{7!}(|x_1|+|x_2|+|\xotau|+|\xttau|)^7
\end{align*}
After substituting the expressions for the upper-bounds of $|\xotau|$ and $|\xttau|$ using Razumikhin's Theorem, the lowest-order term will have $|x_1|^{7/2}=|x_1|^{3.5}$. $\qed$
\end{claim}

\noindent Next, we analyze the higher-order terms and remainder of $\xtdot$. Recall that in the Lyapunov derivative, $\xtdot$ is being multiplied by $x_2^3$. Hence, we require a second-order expansion of $\xtdot$ about $[x_1\ x_2\ \xotau\ \xttau]=\vect{0}$, and we must ensure that the second-order terms are of combined power greater than one.
\begin{claim}{Except for the first-order terms, the second-order expansion of $\xtdot$ in $[x_1\ x_2\ \xotau\ \xttau]$ is of combined power at least 1.5 in $[x_1\ x_2]$.}\\
\noindent\textbf{Proof:} Consider the un-expanded $\xtdot$ and assume that $\ptildetau>0$ in the region where we are considering this function:
\begin{align*}
\xtdot &= 1-(x_2+\hat{s})\frac{\psitau}{\tau}\ptildetau\\
&= 1-\frac{(x_2+\hat{s})}{\tau}\left(c\left(\xttau+\hat{s}-\sqrt[3]{\frac{b(\xotau+\what)}{c}}\right)^3+\xotau+\what-C\tau\right)
\end{align*}
Recall that when expanding this function about $[x_1\ x_2\ \xotau\ \xttau]=\vect{0}$, the zero-order term is zero. 
The second-order terms do not depend on $\xttau$. To see this, let $g=\xtdot$ and consider all second-order partial derivatives of $g$ with respect to $\xttau$:
\begin{align*}
g_{\xttau\xttau} &= -\frac{(x_2+\shat)}{\tau}(6c)\left(\xttau+\hat{s}-\sqrt[3]{\frac{b(\xotau+\what)}{c}}\right) \rightarrow g_{\xttau\xttau}(\vect{0})=0\\
g_{x_2\xttau} &= -\frac{3c}{\tau}\left(\xttau+\hat{s}-\sqrt[3]{\frac{b(\xotau+\what)}{c}}\right)^2 \rightarrow g_{x_2\xttau}(\vect{0})=0\\
g_{\xotau\xttau} &= \frac{(x_2+\shat)}{\tau}(2b)\left(\xttau+\hat{s}-\sqrt[3]{\frac{b(\xotau+\what)}{c}}\right)\left(\frac{b(\xotau+\what)}{c}\right)^{-2/3}
\rightarrow g_{\xotau\xttau}(\vect{0})=0\\
g_{x_1\xttau} &= 0
\end{align*}
Therefore, the second-order terms have combined powers of at least two in $[x_1\ x_2]$. Next, using the Proposition, we can bound the remainder: for some positive constant $M_2$,
\begin{align*}
|R_{\vect{0},2}| \leq \frac{M_2}{3!}(|x_1|+|x_2|+|\xotau|+|\xttau|)^3
\end{align*}
The lowest-order term above is $|x_1|^{3/2}=|x_1|^{1.5}$. $\qed$
\end{claim}

\noindent Finally, we show that the sum of the terms of order four is negative by proving that $Q$ in $\vect{y}^TQ\vect{y}$ is negative definite. 
\begin{align*}
\vect{y}^TQ\vect{y} 
&=\begin{bmatrix}
x_1^2 &
x_1x_2 &
x_2^2
\end{bmatrix}
\begin{bmatrix}
-d_1\alpha & d_1\beta/2 & 0\\[0.3em]
d_1\beta/2 & -d_1\gamma & 0\\[0.3em]
0 & 0 & -d_4/\shat
\end{bmatrix}
\begin{bmatrix}
x_1^2 \\[0.3em]
x_1x_2 \\[0.3em]
x_2^2
\end{bmatrix}
\end{align*}
This first leading principal minor of $Q$, $-d_1\alpha$, is always negative, as needed. The second leading principal minor should be positive:
\begin{align*}
d_1^2\alpha\gamma -d_1^2\frac{\beta^2}{4} &\overset{?}>0\\
\alpha\gamma -\frac{\beta^2}{4} &\overset{?}>0\\
\left(\frac{b^3}{27c^2\hat{s}^7}\right)\left(\frac{b}{\hat{s}^3}\right)-\frac{1}{4}\left(\frac{b^2}{3c\hat{s}^5}\right)^2 &\overset{?}>0\\
\frac{b^4}{27c^2\hat{s}^{10}}-\frac{1}{4}\left(\frac{b^4}{9c^2\hat{s}^{10}}\right) &\overset{?}>0\\
\frac{1}{27}-\frac{1}{36} &\overset{?}>0 \ \checkmark
\end{align*}
The third leading principal minor should be negative:
\begin{align*}
-d_1\alpha\left(d_1d_4\frac{\gamma}{\shat}\right)-d_1\frac{\beta}{2}\left(d_1\frac{\beta}{2}\left(-\frac{d_4}{\shat}\right)\right)&\overset{?}<0\\
-\alpha\left(\frac{\gamma}{\shat}\right)-\frac{\beta}{2}\left(\frac{\beta}{2}\left(-\frac{1}{\shat}\right)\right)&\overset{?}<0\\
-\alpha\gamma+\frac{\beta^2}{4}&\overset{?}<0\\
\alpha\gamma-\frac{\beta^2}{4}&\overset{?}>0
\end{align*}
We see that this condition is equivalent to the previous one (for the second leading principal minor), and hence it is satisfied.

\if{false}
\subsection*{Region of Stability}

For $\dot{V}_2$, we have:
\begin{align*}
\dot{V}_2 &= d_2\left(x_1+\frac{\tau}{\shat^2}x_2\right)\xtdot = -\vect{x}^TQ\vect{x}+d_2\left(x_1+\frac{\tau}{\shat^2}x_2\right)(\shat \dot{x}_{1_{\theta}}+R_2(\vect{h}))
\end{align*}
where $R_2(\vect{h})$ is the remainder of the first-order Taylor series expansion of $\xtdot$ about $\vect{0}$.

For $\dot{V}_1$, we can perform a second-order expansion of $\xodot$ and bound the remainder. We know that the zero, first, and second-order terms are zero in the expansion. Hence, to get the bound on the remainder as in the Proposition, we can compute the third-order partial derivatives of $\xodot$ and bound their absolute values by constants. This is possible, as long as we bound $x_1$ and $\xotau$ in an interval contained by $(\what,\what)$; the reason for this is that some of the partial derivatives contain terms like $\left(\frac{bx_1}{c}+\shat^3\right)^{-l}$, where $l$ is a positive rational number. Hence, we can bound $x_1$ and $\xotau$ in the interval $\left(-\frac{7}{8}\what,\frac{7}{8}\what\right)$, for example, to avoid division by zero. Let $M_1$ denote the maximum absolute value of a third-order partial derivative of $\xodot$. (Note that attempting to obtain this constant is not practical, since it depends on parameter values. For instance, $\left|\frac{\partial \xodot}{\partial x_1^2x_2}\right|$ may yield the maximum for a given system and $\left|\frac{\partial \xodot}{\partial x_1x_2^2}\right|$ may yield the maximum for a system with different parameters.) By the Proposition and Taylor's Theorem, we have
\begin{align*}
|\xodot| &\leq \frac{M_1}{3!}(|x_1|+|x_2|+|\xotau|+|\xttau|)^3
\end{align*}
Using this result, we can bound the value of $\dot{V}_1$:
\begin{align*}
\dot{V}_1 &=(d_1x_1+d_2x_2)\xodot
\leq |(d_1x_1+d_2x_2)||\xodot|\\
\dot{V}_1&\leq \frac{M_1}{6}|(d_1x_1+d_2x_2)|(|x_1|+|x_2|+|\xotau|+|\xttau|)^3
\end{align*}
Note that we now have only quartic terms in the upper-bound for $\dot{V}_1$.

Similarly, let $M_2$ denote the maximum absolute value of a second-order partial derivative of $\xtdot$. By the Proposition, we have:
\begin{align*}
|R_2(\vect{h})| &\leq \frac{M_2}{2}(|x_1|+|x_2|+|\xotau|+|\xttau|)^2
\end{align*}
\fi

Finally, it remains to bound $V$ and $\dot{V}$ with functions $u$, $v$, and $w$ that satisfy all conditions specified in Theorem \ref{thm}.
We note that the arguments above are valid for $|x_1|,\ |x_2|<r$, where $0<r<1$. We keep this in mind for the following bounds.
Recall that previously, we let $d_1=\shat/c$ and $d_4=\tau/\shat$. Then the exact form of the Lyapunov-Razumikhin function is
\begin{align*}
V(\vect{x}) &= \frac{\shat}{2c}x_1^2+\frac{\tau}{4\shat}x_2^4.
\end{align*}
For all bounds, we can use any norm on $\vect{x}$, as long as we are being consistent. We choose the $l_2-$norm. Let 
\begin{align*}
v(||\vect{x}||)=\epsilon_0||\vect{x}||_2^2 = \epsilon_0(x_1^2+x_2^2),
\end{align*}
where $\epsilon_0=\max{\left(\frac{\shat}{2c},\frac{\tau}{4\shat}\right)}$. Then $V(\vect{x})\leq v(||\vect{x}||)$. Also, $v(||\vect{x}||)$ is strictly increasing, $v(||\vect{0}||)=0$, and $v(||\vect{x}||)$ is positive for $||\vect{x}||>0$.
Next, let 
\begin{align*}
u(||\vect{x}||) &=\epsilon_1||\vect{x}||_2^4 = \epsilon_1(x_1^2+x_2^2)^2=\epsilon_1(x_1^4+2x_1^2x_2^2+x_2^4),
\end{align*}
where $\epsilon_1$ is a positive constant of our choice.
We will show that $V(\vect{x}) \geq u(||\vect{x}||)$, or equivalently, $V(\vect{x})-u(||\vect{x}||)\geq 0$ for some choice of $\epsilon_1$.
\begin{align*}
V(\vect{x})-u(||\vect{x}||) &= \frac{\shat}{2c}x_1^2+\frac{\tau}{4\shat}x_2^4-\epsilon_1(x_1^4+2x_1^2x_2^2+x_2^4)\\
&\geq \frac{\shat}{2c}x_1^2+\frac{\tau}{4\shat}x_2^4-\epsilon_1(x_1^2+2x_1^2+x_2^4)\\
&=\left(\frac{\shat}{2c}-3\epsilon_1\right)x_1^2+\left(\frac{\tau}{4\shat}-\epsilon_1\right)x_2^4\\
&\geq 0 \text{ for } \epsilon_1<\min{\left(\frac{\shat}{6c},\frac{\tau}{4\shat}\right)}.
\end{align*}
Also, $u(||\vect{x}||)$ is positive for $||\vect{x}||>0$ and $u(||\vect{0}||)=0$. Hence, condition (\ref{eq:Vbounds}) is satisfied.

So far, we have shown that for a sufficiently small neighborhood of $\vect{x^*}=\vect{0}$, the Lyapunov-Razumikhin candidate in (\ref{eq:lyapfunc}) has a negative definite derivative, \emph{i.e.} $\dot{V}(\vect{x})<0$ for all $\vect{x}\neq \vect{0}$ in this neighborhood and $\dot{V}(\vect{x})=0$ if $\vect{x}=\vect{0}$.
Next, we show that $\dot{V}$ is bounded by a suitable function $w$ as specified in Theorem \ref{thm}.
Recall that we have shown that
\begin{align*}
\dot{V} &= d_1\left(-\alpha x_1^4+\beta x_1^3x_2-\gamma x_1^2x_2^2\right)-\frac{d_4}{\hat{s}}x_2^4 +h.o.t.
\end{align*}
We can express the lower-order terms in matrix form, as follows:
\begin{align*}
\dot{V} &=
-\begin{bmatrix}
x_1^2 &
\sqrt{2}x_1x_2 &
x_2^2
\end{bmatrix}
\begin{bmatrix}
d_1\alpha & -\frac{d_1\beta}{2\sqrt{2}} & 0\\[0.3em]
-\frac{d_1\beta}{2\sqrt{2}} & \frac{d_1\gamma}{2} & 0\\[0.3em]
0 & 0 & d_4/\shat
\end{bmatrix}
\begin{bmatrix}
x_1^2 \\[0.3em]
\sqrt{2}x_1x_2 \\[0.3em]
x_2^2
\end{bmatrix}
  +h.o.t.
\end{align*}
Let's call the matrix above $\tilde{Q}$. Since we have previously shown that the sum of the lower-order terms in $\dot{V}$ is a negative definite function, it follows that $\tilde{Q}$ is positive definite, so all of its eigenvalues are strictly positive. Since $\tilde{Q}$ is a real and symmetric matrix, its Rayleigh quotient is bounded below by $\lambda_{\min}[\tilde{Q}]$. Hence, 
\begin{align*}
\dot{V} &\leq -\lambda_{\min}[\tilde{Q}] \left|\left|\begin{bmatrix}
x_1^2 \\[0.3em]
\sqrt{2}x_1x_2 \\[0.3em]
x_2^2
\end{bmatrix}\right|\right|_2^2+h.o.t.\\
&=  -\lambda_{\min}[\tilde{Q}] (x_1^4+2x_1^2x_2^2+x_2^4)+h.o.t.\\
&= -\lambda_{\min}[\tilde{Q}] ||\vect{x}||_2^4+h.o.t.\\
\end{align*}
Previously, we showed that the higher-order terms have orders of at least $4.5$. Then for $|x_1|,$ $|x_2|$ small enough, there exists a positive constant $K$ such that
\begin{align*}
h.o.t. \leq K||\vect{x}||_2^4.
\end{align*}
\begin{claim}
\label{claim:limK}
\begin{align*}
\text{Let }f(x_1,x_2) &=\frac{h.o.t.}{||\vect{x}||_2^4}.\text{ Then }
\lim\limits_{(x_1,x_2)\to (0,0)}\frac{h.o.t.}{||\vect{x}||_2^4}=0.
\end{align*}
\noindent\textbf{Proof:} To prove our claim, we will show that for every $\epsilon>0$, there exists a $\delta >0$ so that whenever $0<\sqrt{x_1^2+x_2^2}<\delta$, $|f(x_1,x_2)|<\epsilon$.
The higher-order terms are a sum of terms that have format $c_0x_1^{c_1}x_2^{c_2}$, where $c_0$ is either a constant or a function of $x_{1_{\theta+\tau}}$ and $x_{2_{\theta+\tau}}$, as in the case of the higher-order terms that arise from the $x_2^3\dot{x}_{1_{\theta}}\tau$ term of $\dot{V}$. We showed that in all cases, $|c_0|$ is bounded above by a positive constant. In addition, we showed that $c_1+c_2\geq 4.5$.
Let $n$ be the number of higher-order terms (note: $n$ is finite). Then we can write $|f(x_1,x_2)|$ as follows:
\begin{align*}
|f(x_1,x_2)| &=\frac{1}{||\vect{x}||_2^4}\left|\sum\limits_{i=1}^n c_0^{(i)}x_1^{c_1^{(i)}}x_2^{c_2^{(i)}}\right|,
\end{align*}
where the superscript $(i)$ corresponds to the $i$th higher-order term. Clearly,
\begin{align*}
|f(x_1,x_2)| &\leq \frac{1}{||\vect{x}||_2^4}\sum\limits_{i=1}^n \left|c_0^{(i)}x_1^{c_1^{(i)}}x_2^{c_2^{(i)}}\right|.
\end{align*}
Consider any term $|c_0x_1^{c_1}x_2^{c_2}|$. Factor out any combination $|x_1|^a|x_2|^b$ such that $a+b=4$. We know that 
\begin{align*}
|x_1|,|x_2|\leq \sqrt{x_1^2+x_2^2}.
\end{align*}
Then 
\begin{align*}
|x_1|^a|x_2|^b &\leq \left(\sqrt{x_1^2+x_2^2}\right)^4 = ||\vect{x}||_2^4,\\
|c_0x_1^{c_1}x_2^{c_2}| &\leq |c_0g(x_1,x_2)|||\vect{x}||_2^4
\end{align*}
where $g(x_1,x_2)$ is defined s.t. $g(x_1,x_2)x_1^ax_2^b=x_1^{c_1}x_2^{c_2}$. Hence,
\begin{align*}
|f(x_1,x_2)| &\leq \frac{1}{||\vect{x}||_2^4}\sum\limits_{i=1}^n \left|c_0^{(i)}g^{(i)}(x_1,x_2)\right|||\vect{x}||_2^4 = \sum\limits_{i=1}^n \left|c_0^{(i)}g^{(i)}(x_1,x_2)\right|.
\end{align*}
Each function $g(x_1,x_2)$ necessarily has the form
\begin{align*}
g(x_1,x_2) &= x_1^{l_1}x_2^{l_2},\ l_1,l_2\geq 0,\ l_1+l_2\geq \frac{1}{2}.
\end{align*}
Hence, we can bound the absolute value of each of these functions by a function of $\delta$.
\begin{align*}
|x_{1/2}| &\leq \sqrt{x_1^2+x_2^2}<\delta,\\
|x_{1/2}|^{l_{1/2}} &\leq \left(\sqrt{x_1^2+x_2^2}\right)^{l_{1/2}} <\delta^{l_{1/2}},\\
|g(x_1,x_2)| &\leq \left(\sqrt{x_1^2+x_2^2}\right)^{l_1+l_2} < \delta^{l_1+l_2}.
\end{align*}
This gives us
\begin{align*}
|f(x_1,x_2)| &\leq \sum\limits_{i=1}^n \left|c_0^{(i)}\right|\delta^{l_1^{(i)}+l_2^{(i)}}.
\end{align*}
We would like the sum above to be less than a given $\epsilon>0$. We can always find a $\delta>0$ small enough to make this happen. \qed
\end{claim}
By Claim \ref{claim:limK}, we can always find a $K$ small enough by restricting $x_1$ and $x_2$ into a smaller neighborhood around $\vect{0}$. Hence, we can find a $K<\lambda_{\min}[\tilde{Q}]$, which would give us
\begin{align*}
\dot{V} &\leq -(\lambda_{\min}[\tilde{Q}]-K) ||\vect{x}||_2^4
\end{align*}
where $\lambda_{\min}[\tilde{Q}]-K>0$.
By inspection, we have found a $w(||\vect{x}||)$ that satisfies condition (\ref{eq:Vdotw}) under (\ref{eq:Vdotcondstr}). In addition, 
$w(||\vect{x}||))>0$ when $||\vect{x}||>0$, as necessary. Finally, $\lim_{||\vect{x}||\to\infty}u(||\vect{x}||)=\infty$. By Theorem \ref{thm}, we have shown that the function that drives CUBIC's \cwnd, (\ref{eq:cubiccwnd}), is locally uniformly asymptotically stable.
\subsubsection*{Convergence}
Using the Lyapunov-Razumikhin function and its derivative, it is possible to explicitly demonstrate the convergence of trajectories to the fixed point. In the analysis below, we assume that $t_0=0$.
Recall that 
\begin{align*}
V(\vect{x}) &\leq \epsilon_0||\vect{x}||_2^2\\
\rightarrow V^2(\vect{x}) &\leq \epsilon_0^2||\vect{x}||_2^4.
\end{align*}
This gives us
\begin{align*}
\dot{V} &\leq -\frac{(\lambda_{\min}[\tilde{Q}]-K)}{\epsilon_0^2}V^2\\
\rightarrow\frac{\dot{V}}{V^2} &\leq -\frac{(\lambda_{\min}[\tilde{Q}]-K)}{\epsilon_0^2}.
\end{align*}
We note that 
\begin{align*}
\frac{d}{dt}\left(-\frac{1}{V}\right)&=\frac{\dot{V}}{V^2}\leq -\frac{(\lambda_{\min}[\tilde{Q}]-K)}{\epsilon_0^2}, \text{ so}\\
\frac{d}{dt}\left(\frac{1}{V}\right)&\geq\frac{(\lambda_{\min}[\tilde{Q}]-K)}{\epsilon_0^2}.
\end{align*}
Then, solving the differential inequality,
\begin{align*}
\int_0^t \frac{d}{ds}\left(\frac{1}{V}\right)ds = \frac{1}{V(t)}-\frac{1}{V(0)} &\geq \int_0^t\frac{(\lambda_{\min}[\tilde{Q}]-K)}{\epsilon_0^2}ds=\frac{(\lambda_{\min}[\tilde{Q}]-K)}{\epsilon_0^2}t,\\
\frac{1}{V(t)} &\geq \frac{(\lambda_{\min}[\tilde{Q}]-K)}{\epsilon_0^2}t+\frac{1}{V(0)},\\
V(t) &\leq \frac{1}{\frac{(\lambda_{\min}[\tilde{Q}]-K)}{\epsilon_0^2}t+\frac{1}{V(0)}}.
\end{align*}
Since $V(t)\geq \epsilon_1||\vect{x}||_2^4$,
\begin{align}
||\vect{x}||_2^4 \leq \frac{1}{\frac{\epsilon_1(\lambda_{\min}[\tilde{Q}]-K)}{\epsilon_0^2}t+\frac{\epsilon_1}{V(0)}}.
\label{eq:xbound}
\end{align}
\if{false}
Therefore,
\begin{align*}
\dot{V} &\leq -\frac{\lambda_{\min}[\tilde{Q}]}{\epsilon_0^2}V^2+K(|x_1|^{1/4}+|x_2|^{1/4})||\vect{x}||_2^4\\
\end{align*}
Recall from the previous section that $V(\vect{x})\geq \epsilon_1||\vect{x}||_2^4$. Hence,
\begin{align*}
\dot{V} &\leq -\frac{\lambda_{\min}[\tilde{Q}]}{\epsilon_0^2}V^2+K(|x_1|^{1/4}+|x_2|^{1/4})\frac{V}{\epsilon_1}
\end{align*}
For $|x_1|^{1/4}+|x_2|^{1/4}\leq \frac{1}{K}$,
\begin{align*}
\dot{V} &\leq -\frac{\lambda_{\min}[\tilde{Q}]}{\epsilon_0^2}V^2+\frac{V}{\epsilon_1}
\end{align*}
Above, we have a Bernoulli differential inequality, which can be solved using variable substitution and the method of integrating factors. First, we write it in standard form:
\begin{align*}
\dot{V} -\frac{V}{\epsilon_1}&\leq -\frac{\lambda_{\min}[\tilde{Q}]}{\epsilon_0^2}V^2
\end{align*}
Divide both sides by $V^2$:
\begin{align*}
\frac{\dot{V}}{V^2} -\frac{1}{\epsilon_1V}&\leq -\frac{\lambda_{\min}[\tilde{Q}]}{\epsilon_0^2}
\end{align*}
Let
\begin{align*}
w=\frac{1}{V} \rightarrow \dot{w} = -\frac{\dot{V}}{V^2}
\end{align*}
Substitute:
\begin{align*}
-\dot{w} -\frac{1}{\epsilon_1}w&\leq -\frac{\lambda_{\min}[\tilde{Q}]}{\epsilon_0^2}\\
\dot{w} +\frac{1}{\epsilon_1}w&\geq \frac{\lambda_{\min}[\tilde{Q}]}{\epsilon_0^2}
\end{align*}
Let the integrating factor be
\begin{align*}
M(s) &= e^{\int \frac{1}{\epsilon_1}ds}=e^{s/\epsilon_1}
\end{align*}
Multiply both sides of the inequality by $M(s)$:
\begin{align*}
e^{s/\epsilon_1}\left(\dot{w} +\frac{1}{\epsilon_1}w\right)&\geq \frac{\lambda_{\min}[\tilde{Q}]}{\epsilon_0^2}e^{s/\epsilon_1}
\end{align*}
We note that 
\begin{align*}
\frac{d}{ds}\left(w(s)e^{s/\epsilon_1}\right) &= e^{s/\epsilon_1}\left(\dot{w}(s) +\frac{1}{\epsilon_1}w(s)\right)
\end{align*}
Hence,
\begin{align*}
\frac{d}{ds}\left(w(s)e^{s/\epsilon_1}\right)&\geq \frac{\lambda_{\min}[\tilde{Q}]}{\epsilon_0^2}e^{s/\epsilon_1}
\end{align*}
Taking the definite integral of both sides, we have
\begin{align}
\int\limits_{0}^t\frac{d}{ds}\left(w(s)e^{s/\epsilon_1}\right)ds&\geq \int\limits_{0}^t\frac{\lambda_{\min}[\tilde{Q}]}{\epsilon_0^2}e^{s/\epsilon_1}ds\label{eq:int}
\end{align}
By the fundamental theorem of calculus,
\begin{align*}
\int\limits_{0}^t\frac{d}{ds}\left(w(s)e^{s/\epsilon_1}\right)ds &= w(t)e^{t/\epsilon_1}-w(0)
\end{align*}
Substituting this into (\ref{eq:int}) and evaluating the r.h.s. of (\ref{eq:int}), we have:
\begin{align*}
w(t)e^{t/\epsilon_1}-w(0) &\geq \frac{\lambda_{\min}[\tilde{Q}]\epsilon_1}{\epsilon_0^2}\left(e^{t/\epsilon_1}-1\right)\\
w(t)-w(0)e^{-t/\epsilon_1} &\geq \frac{\lambda_{\min}[\tilde{Q}]\epsilon_1}{\epsilon_0^2}\left(1-e^{-t/\epsilon_1}\right)\\
\frac{1}{V(t)}-\frac{1}{V(0)}e^{-t/\epsilon_1} &\geq \frac{\lambda_{\min}[\tilde{Q}]\epsilon_1}{\epsilon_0^2}\left(1-e^{-t/\epsilon_1}\right)\\
\frac{1}{V(t)} &\geq \frac{\lambda_{\min}[\tilde{Q}]\epsilon_1}{\epsilon_0^2}\left(1-e^{-t/\epsilon_1}\right)+\frac{1}{V(0)}e^{-t/\epsilon_1}\\
V(t) &\leq \frac{1}{\frac{\lambda_{\min}[\tilde{Q}]\epsilon_1}{\epsilon_0^2}\left(1-e^{-t/\epsilon_1}\right)+\frac{1}{V(0)}e^{-t/\epsilon_1}}
\end{align*}
Since we know that $V(\vect{x}) \geq \epsilon_1 ||\vect{x}||_2^4$, we have
\begin{align}
\epsilon_1 ||\vect{x}||_2^4 &\leq \frac{1}{\frac{\lambda_{\min}[\tilde{Q}]\epsilon_1}{\epsilon_0^2}\left(1-e^{-t/\epsilon_1}\right)+\frac{1}{V(0)}e^{-t/\epsilon_1}}\nonumber\\
\rightarrow  ||\vect{x}||_2^4 &\leq \frac{1}{\frac{\lambda_{\min}[\tilde{Q}]\epsilon_1^2}{\epsilon_0^2}\left(1-e^{-t/\epsilon_1}\right)+\frac{\epsilon_1}{V(0)}e^{-t/\epsilon_1}}
\label{eq:xbound}
\end{align}
\fi
We can simplify this bound using the definition of uniform stability:
\begin{definition}[Definition 1.1 from Stability of Time-Delay Systems]
For the system described by $\dot{x}=f(t,x_t)$, the trivial solution $x(t)=0$ is said to be stable if for any $t_0\in\mathbb{R}$ and any $\epsilon>0$, there exists a $\delta(t_0,\epsilon)>0$ such that $||x_{t_0}||_c<\delta$ implies $||x(t)||<\epsilon$ for $t>t_0$. It is said to be uniformly stable if it is stable and $\delta(t_0,\epsilon)$ can be chosen independently of $t_0$. It is uniformly asymptotically stable if it is uniformly stable and there exists a $\delta_a>0$ such that for any $\eta>0$, there exists a $T(\delta_a,\eta)$, such that $||x_{t_0}||_c<\delta_a$ implies $||x(t)||<\eta$ for $t\geq t_0+T$ and $t_0\in\mathbb{R}$.
\end{definition}
Above, $||\phi||_c=\max_{a\leq\xi\leq b}||\phi(\xi)||$ for $\phi\in\mathcal{C}[a,b]$ and $x_{t_0}=\phi$ or $x(t_0+\theta)=\phi(\theta)$, $-\tau\leq\theta\leq 0$. 
From (\ref{eq:xbound}), we have that
\begin{align*}
||\vect{x}||_2^4 \leq \frac{1}{\frac{\epsilon_1(\lambda_{\min}[\tilde{Q}]-K)}{\epsilon_0^2}t+\frac{\epsilon_1}{V(0)}} 
\leq \frac{1}{\frac{\epsilon_1}{V(0)}} = \frac{V(0)}{\epsilon_1} \leq \frac{\epsilon_0||\vect{x}(t_0)||_2^2}{\epsilon_1} < \frac{\epsilon_0\delta^2}{\epsilon_1}.
\end{align*}
We would like
\begin{align*}
||\vect{x}||_2 &<\epsilon, \text{ or}\\
||\vect{x}||_2^4 &< \epsilon^4.
\end{align*}
So let 
\begin{align*}
\frac{\epsilon_0\delta^2}{\epsilon_1} &<\epsilon^4\\
\rightarrow\delta &<\epsilon^2\sqrt{\frac{\epsilon_1}{\epsilon_0}}.
\end{align*}
This bound on $\delta$ provides a measure on the basin of attraction of the fixed point of a system using the system's parameters. \emph{I.e.}, it indicates how close the initial conditions must be to the fixed point in order to guarantee stability. One possible way to apply this bound is in the implementation of TCP: one could specify the initial slow start threshold to be close to $\what$, so that when CUBIC's congestion avoidance phase begins, the systems is more likely to settle into its stable state.
\if{false}
\begin{align*}
\dot{V} &\leq -\frac{V^2}{\epsilon_0^2}\left(\lambda_{\min}[\tilde{Q}]-\frac{\lambda_{\min}[\tilde{Q}]}{2}\right)\\
\rightarrow \dot{V} &\leq -\frac{V^2}{\epsilon_0^2}\frac{\lambda_{\min}[\tilde{Q}]}{2}
\end{align*}
Now, we can solve this differential inequality:
\begin{align}
\dot{V} &\leq -kV^2, \text{ where } k=\frac{\lambda_{\min}[\tilde{Q}]}{2\epsilon_0^2}\nonumber\\
\rightarrow \frac{\dot{V}}{V^2} &\leq -k\nonumber\\
\frac{d}{dt}\left(-\frac{1}{V}\right) &= \frac{\dot{V}}{V^2} \leq -k \label{eq:Vderivineq}
\end{align}
Using the fundamental theorem of calculus,
\begin{align*}
\int\limits_{0}^{t}\frac{d}{ds}\left(-\frac{1}{V(s)}\right)ds &= -\frac{1}{V(t)}-\left(-\frac{1}{V(0)}\right)
\end{align*}
We can rewrite the above as:
\begin{align*}
\frac{1}{V(t)} &= \frac{1}{V(0)} +\int\limits_{0}^{t}\frac{d}{ds}\left(\frac{1}{V(s)}\right)ds
\end{align*}
From (\ref{eq:Vderivineq}), we know that $\frac{d}{ds}\left(\frac{1}{V(s)}\right) \geq k$, so
\begin{align*}
\frac{1}{V(t)} &\geq \frac{1}{V(0)} +\int\limits_{0}^{t}kds\\
\frac{1}{V(t)} &\geq \frac{1}{V(0)} +kt\\
V(t) &\leq \frac{1}{\frac{1}{V(0)} +kt}
\end{align*}
Since we know that $V(\vect{x}) \geq \epsilon_1 ||\vect{x}||_2^4$, we have
\begin{align*}
\epsilon_1 ||\vect{x}||_2^4 &\leq \frac{1}{\frac{1}{V(0)} +kt}\\
\rightarrow  ||\vect{x}||_2^4 &\leq \frac{1}{\epsilon_1\left(\frac{1}{V(0)} +\frac{\lambda_{\min}[\tilde{Q}]}{2\epsilon_0^2}t\right)}
\end{align*}
\fi
\subsection*{Summary}
For the system described by (\ref{sys:model}), the following properties hold:
\begin{description}
\item[(a)] The system has a unique fixed point $\vect{x^*}=\vect{0}$.
\item[(b)] The system has a unique solution in a neighborhood of this fixed point.
\item[(c)] The fixed point is locally uniformly asymptotically stable for $\vect{x}$ small enough and in addition,
\begin{description}
\item[(i)] $x_1$ and $\xotau$ are constrained to $[-\rho\what,\rho\what]$, $0<\rho<1$.
\item[(ii)] $|x_1|,|x_2|<1$.
\end{description}
\item[(d)] The solution is bounded according to (\ref{eq:xbound}) for $|x_1|$ and $|x_2|$ small enough.
\end{description} 
\section{Simulations}
\label{simulations}
\begin{figure*}[t!]
\centering
\subfloat[$\tau=1$ms, 1 flow]{{\includegraphics[width=0.49\textwidth]{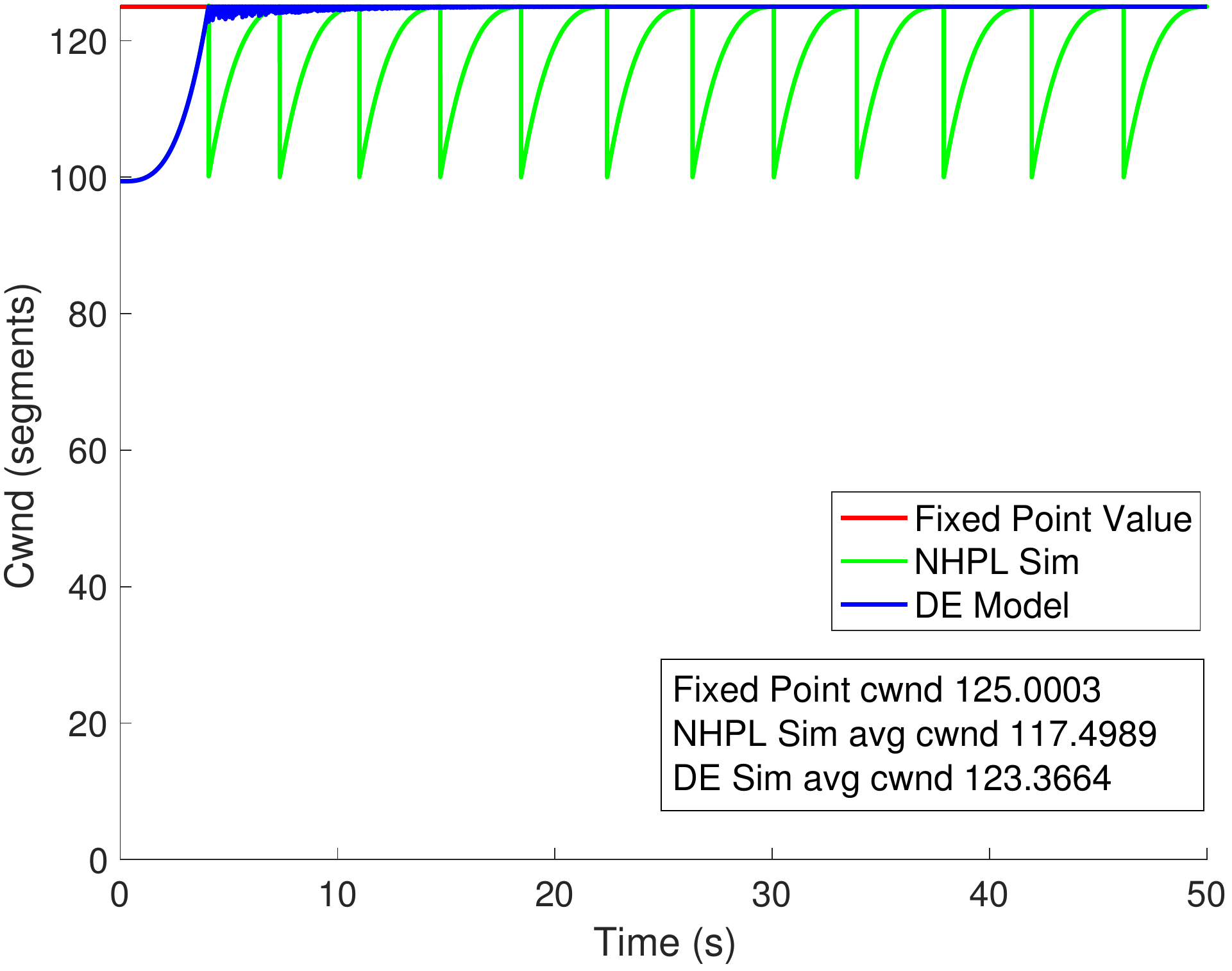}}\label{fig:NHPLvsDEa}}
\subfloat[$\tau=1$ms, 20 flows]{{\includegraphics[width=0.49\textwidth]{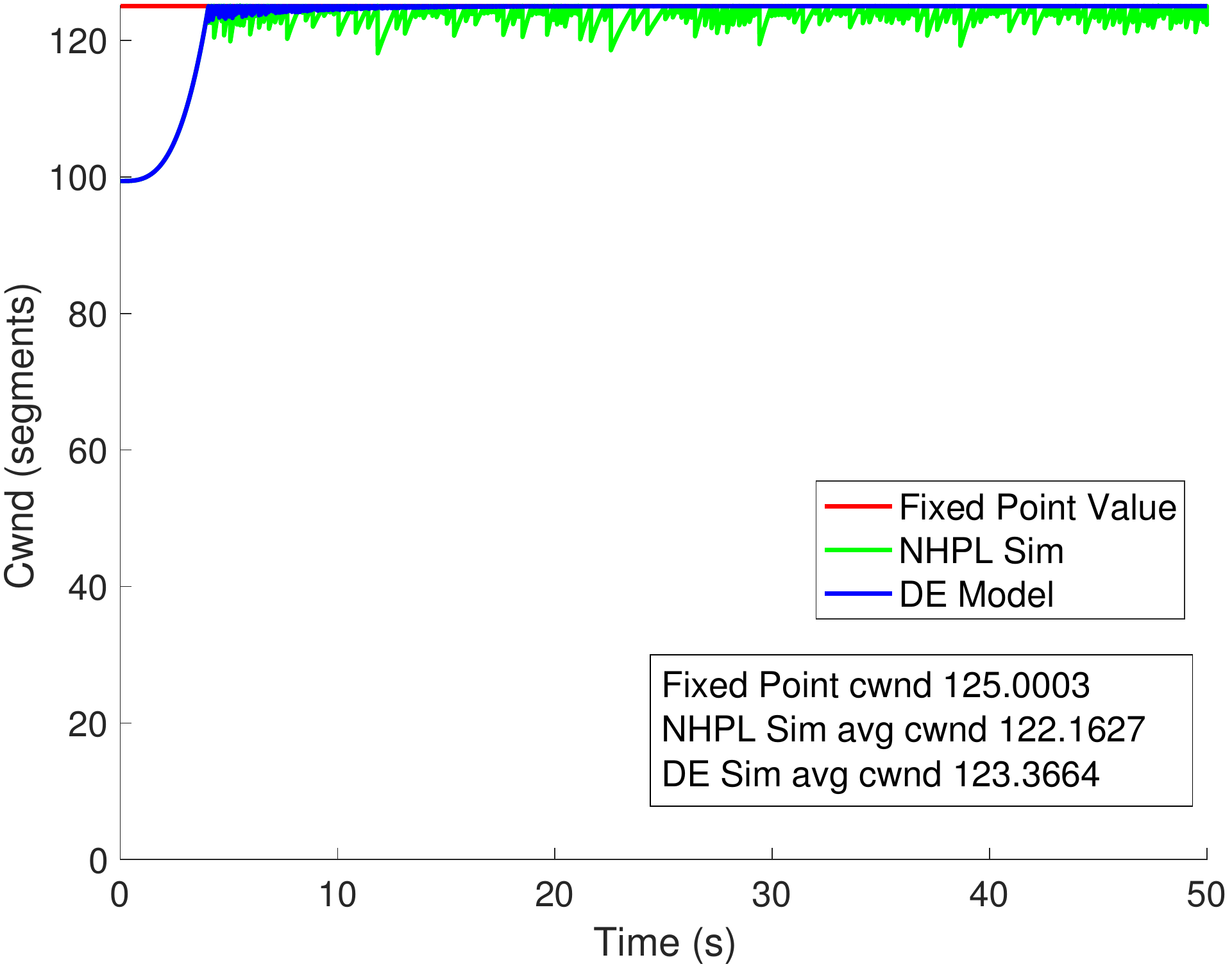}}\label{fig:NHPLvsDEb}}\\
\subfloat[$\tau=10$ms, 1 flow]{{\includegraphics[width=0.49\textwidth]{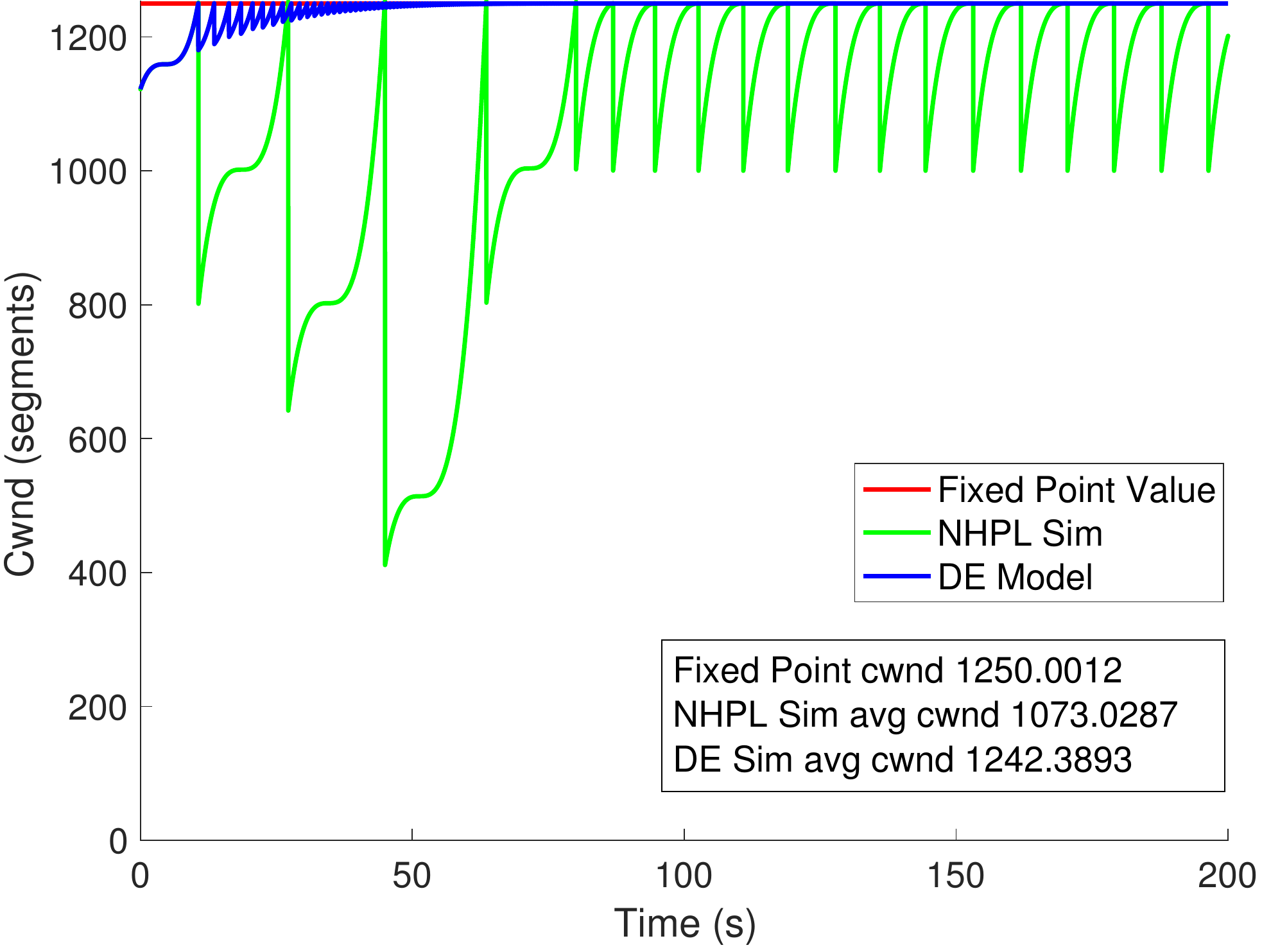} }\label{fig:NHPLvsDEc}}
\subfloat[$\tau=10$ms, 20 flows]{{\includegraphics[width=0.49\textwidth]{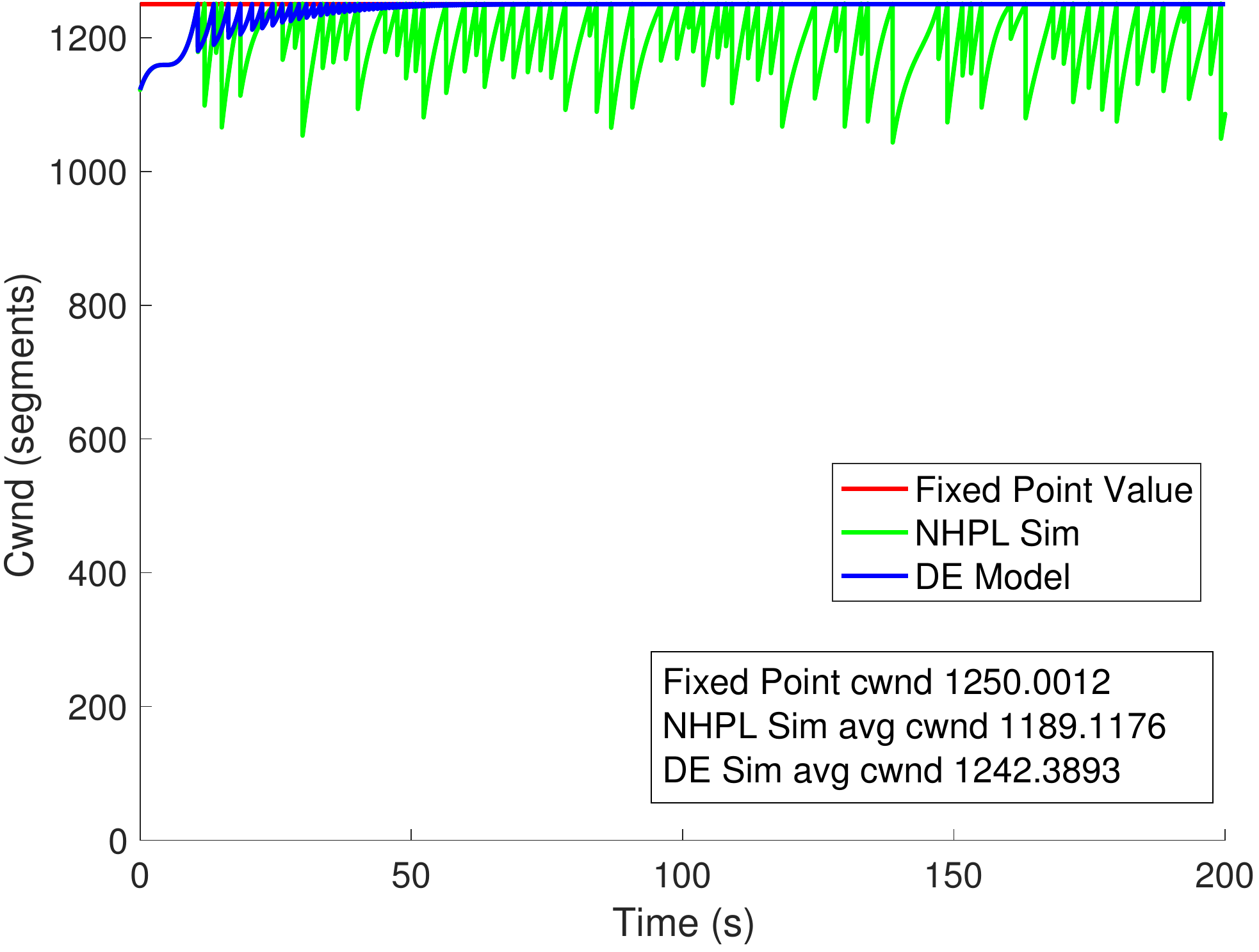} }\label{fig:NHPLvsDEd}}
\caption{Comparison of average \cwnd (computed post-transient phase) generated by NHPL simulations against steady-state \cwnd generated by model (\ref{sys:model}) for TCP CUBIC. Also shown is the fixed-point value of \cwnd. Per-flow capacity $C=1$ Gbps.}
\label{fig:NHPLvsDE}
\end{figure*}

\begin{figure}[t!]
\centering
\subfloat[$\tau=1$ms, stable]{{\includegraphics[width=0.49\textwidth]{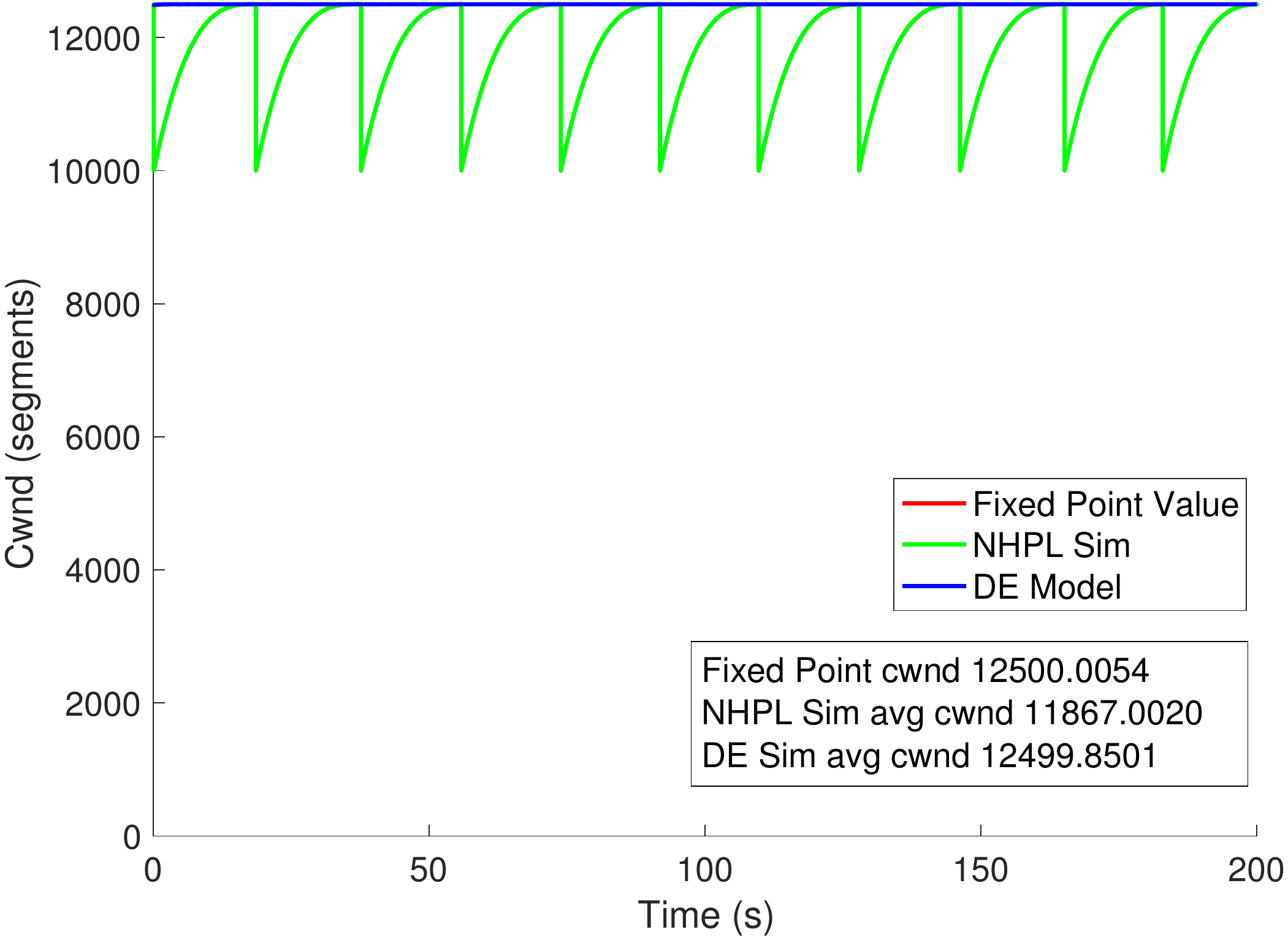}}\label{fig:bigrttstable}}
\subfloat[$\tau=1$ms, unstable]{{\includegraphics[width=0.49\textwidth]{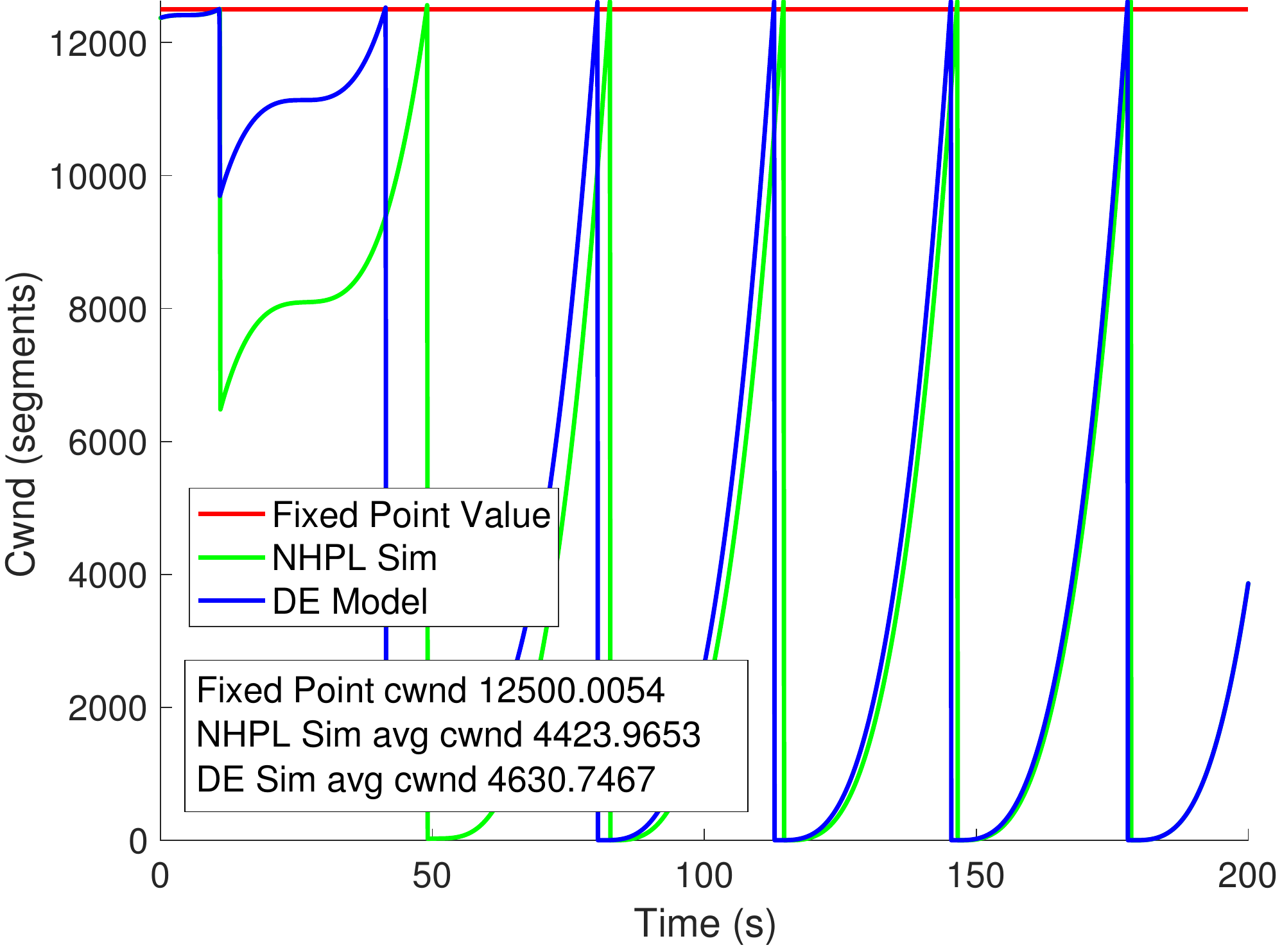}}\label{fig:bigrttunstable}}
\caption{The impact of initial conditions on stability. For both (a) and (b), $C=1$ Gbps, $\tau=100$ms. In (a), there is one flow whose initial conditions $W(0)$ and $s(0)$ are very close to the fixed point values $\what$ and $\shat$, respectively. Both the NHPL simulation and the model exhibit stability. In (b), there are 20 flows whose initial conditions are set too far from the fixed point values, destabilizing the flows in both the NHPL simulation and the DE system.}
\label{fig:stabunstab}
\end{figure}

\begin{figure}[t!]
\centering
\includegraphics[width=0.55\textwidth]{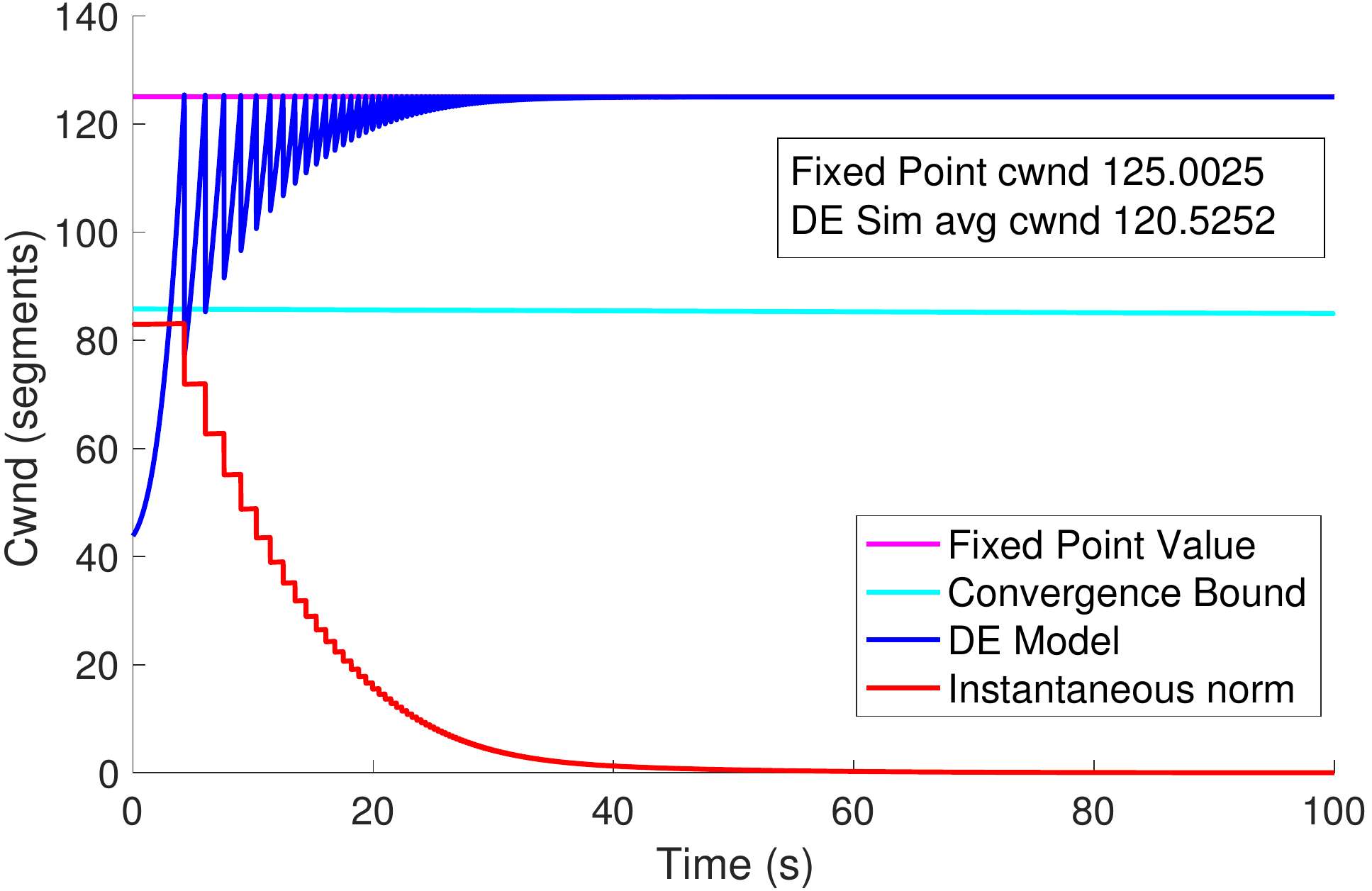}
\caption{Convergence for CUBIC. At the top is the \cwnd generated by DEs as it converges to the fixed point value of \cwnd. Below these two curves is a comparison of $||\vect{x}||_2$ against the analytical bound in (\ref{eq:xbound}). ${C=100}$ Mbps, ${\tau=10}$ms. 
}
\label{fig:convergence}
\end{figure}
We use simulations to validate model (\ref{sys:model}) and its stability analysis for TCP CUBIC.
Our simulation framework treats loss as a non-homogenous Poisson process and generates new loss events based on a user-defined probability of loss model. 
A detailed description of the framework is provided in the Appendix.
An advantage  of using this framework for validating the DEs over, for example, NS3, is that we can observe the behavior of solely the congestion avoidance phase of an algorithm, which allows us to more easily verify the theoretical analysis of the controller's stability. Moreover, as we observe from simulations of the DEs, an algorithm's stability can be highly sensitive to the initial conditions specified at the beginning of the congestion avoidance phase. The initial conditions are values of $\wmax(0)$ and $s(0)$ for all flows, and we can control them more easily with our simulation framework. This can be especially useful when testing the region of stability for a given system.

Figure \ref{fig:NHPLvsDE} compares the average \cwnd generated by the Non-Homogeneous Poisson Loss (NHPL) simulations against the average value of \cwnd generated by the DEs. The fixed-point value of \cwnd, $\what$, is also shown (albeit sometimes entirely hidden by the DE curve because of fast convergence). 
All flows in this figure have a per-flow capacity of 1 Gbps, while the round-trip time is varied (these combinations of $C$ and $\tau$ are sufficient to generate a diverse set of behaviors). All flows have $b=0.2$ and $c=0.4$ (the default values used in Linux implementations of CUBIC). 

Panel \ref{fig:NHPLvsDE}(a) shows a single stable flow with $\tau=1$ms. The transient response of both simulations is clearly visible, and we observe that they reach steady-state within a similar period. Not shown in this panel is the value of $\hat{s}\approx4$s. By observing the time between losses in the NHPL simulation, we see that there is a close agreement. Panel \ref{fig:NHPLvsDE}(b) shows the same experiment, but with 20 flows. As expected, the average value of \cwnd from the NHPL simulation approaches $\what$ as the number of flows is increased. Panels \ref{fig:NHPLvsDE}(c) and \ref{fig:NHPLvsDE}(d) show one and 20 flows, respectively, for $\tau=10$ms. The initial conditions (values of $s(0)$ and $\wmax(0)$) are deliberately far enough from the fixed point to demonstrate a more dramatic transient response from both simulations. Figure \ref{fig:stabunstab} shows two examples of 100ms flows: in (a), there is a single flow that is stable, while the initial conditions in (b) cause instability for 20 flows in both the DEs and NHPL simulation.

Figure \ref{fig:convergence} illustrates the transient and steady-state responses of a flow with $C=100$ Mbps and $\tau=10$ms, as well $||\vect{x}||_2$ as it compares to the convergence bound (\ref{eq:xbound}). Observe that $||\vect{x}||_2$ is always below the bound and approaches zero as the flow reaches steady state. The bound appears flat in this example because for this system, $V(t_0)$ dominates in the denominator. We observe this phenomenon for many systems; this implies that the initial conditions are crucial for a flow's stability.

\if{false}
Figure \ref{fig:convergence} presents an example demonstrating CUBIC's convergence to the fixed point $\vect{x^*}$. Each panel shows the value of the fixed point, the response of the system, bounds (\ref{eq:xbound}) and (\ref{eq:bound2}), and the instantaneous norm of $\vect{x}(t)$. Note that since we can only claim local stability, the convergence result is valid only in a neighborhood of $\vect{x^*}$. Hence, if we choose $|x_1(0)|$ and $|x_2(0)|$ small enough, then $||\vect{x}(t)||_2$ remains within both bounds, as in panel (\ref{fig:conva}) and eventually decays to zero. Consider what happens if the initial values are too far from $\vect{x^*}$, as in panel (\ref{fig:convb}). Initially, $||\vect{x}(t)||_2$ ventures outside of both bounds, but eventually converges to zero. Although our analysis does not guarantee convergence in such a case, this example illustrates that at least for some systems, the basin of attraction is larger than the one specified by our Lyapunov function.
\fi

\section{Conclusion}
\label{conclusion}
The main contribution of this work is a novel and versatile fluid model for \cwnd- and rate-based data transport algorithms. 
The model is structured so that the differential equations are not dependent on the specific window or rate function of a controller. 
As a result, this framework  
offers opportunities to model and analyze the stability of a diverse set of controllers whose window or rate functions may not be linear and whose increase and decrease rules may not be given in explicit form.
We apply this model to two different algorithms: TCP Reno and CUBIC. For the former, we prove that the new model is equivalent to a well-established model for Reno. For the latter, the new model succeeds where traditional methods of modeling \cwnd are ineffective. We go on to analyze the fluid model for CUBIC and discover that for a given probability of loss model, its window is locally uniformly asymptotically stable. We derive a convergence bound on the solution of the system as a function of the system parameters.  Simulations of the model support our theoretical results. As a future direction, we plan to validate the model against a packet-based simulation, as well as analyze the model using alternate loss probability functions.
\section{Appendix}
We introduce a method of simulating the evolution of a congestion window given $W(t)$ -- \cwnd as a function of time, and $\lambda(t)$ -- loss rate as a function of time. We first describe the procedure for generating loss events given arbitrary $W(t)$ and $\lambda(t)$. We then consider a specific loss model and discuss the workarounds necessary when dealing with capacity constraints and time delays. The final result is an algorithm whose pseudocode we present in detail. Finally, we illustrate the operation of the algorithm using an example \cwnd trajectory.

\subsection{Generating Loss Events}
We would like to generate inter-loss times given a loss rate function $\lambda(t)$. In order to do so, we apply the Inverse Transform Method on the Poisson distribution, described in the following proposition.

\begin{proposition}
\label{prop:1.1}
Suppose a loss event occurs at time $t_0$.  The time to the next loss is given by $T$ where
\[ \int_{t_0}^{t_0+T}{\lambda(t) dt} = -\ln{u},\]
where $u$ is randomly generated from the uniform distribution $U(0, 1)$.
\end{proposition}
{\bf Proof.}
Note that $\lambda(t)$ denotes a Non-Homogeneous Poisson Process, where the number of events
between $s$ and $t, N_s(t)$ has a Poisson distribution with parameter $m_s(t) = \int_{s}^{t}{\lambda(\tau) d\tau}$,
\[P(N_s(t) = k) = \frac{m_s(t)^k}{k!}e^{-m_s(t)}.\]

We can then write the CDF of the time from $t_0$ to the next loss as
\begin{align*}
F_{X_{t_0}}(T) &= 1-P(N_{t_0}(T) = 0) = P(N_{t_0}(T) > 0) \\
&= 1 - \exp{\Big( - \int_{t_0}^{t_0+T}{\lambda(t) dt} \Big)}.
\end{align*}

Note that a CDF can be seen as a random variable with uniform distribution $U(0,1)$, and
can be sampled by generating uniform random numbers (this is known as Inverse Transform Sampling). Therefore, inter-loss time samples
can be generated as $T = F_{X_{t_0}}^{-1}(u)$. From the above equation we obtain

\[\int_{t_0}^{t_0+T}{\lambda(t) dt} = -\ln{(1-u)} \equiv -\ln{u},\]
where the last equivalence follows from the fact that if $u$ is uniformly distributed
between 0 and 1, so is $1-u$.
\done
\subsection{Delays and Capacity Constraints}
In TCP (and most other data transport protocols), the loss rate is a function of the sending rate $W(t)/\tau$ and of a probability of loss model $p(t)$:
\begin{align}
\lambda(t) = \frac{W(t)p(t)}{\tau}.
\label{eq:lambda}
\end{align}
Therefore, in order to obtain a sample of the time until next loss, the following equation can be solved for $T$:
\begin{align}
\frac{1}{\tau}\int_{t_0}^{t_0+T}W(t)p(t)dt = -\ln(u).
\label{eq:solveForT}
\end{align}

\begin{figure*}[ht]
\centering
\includegraphics[width=\textwidth]{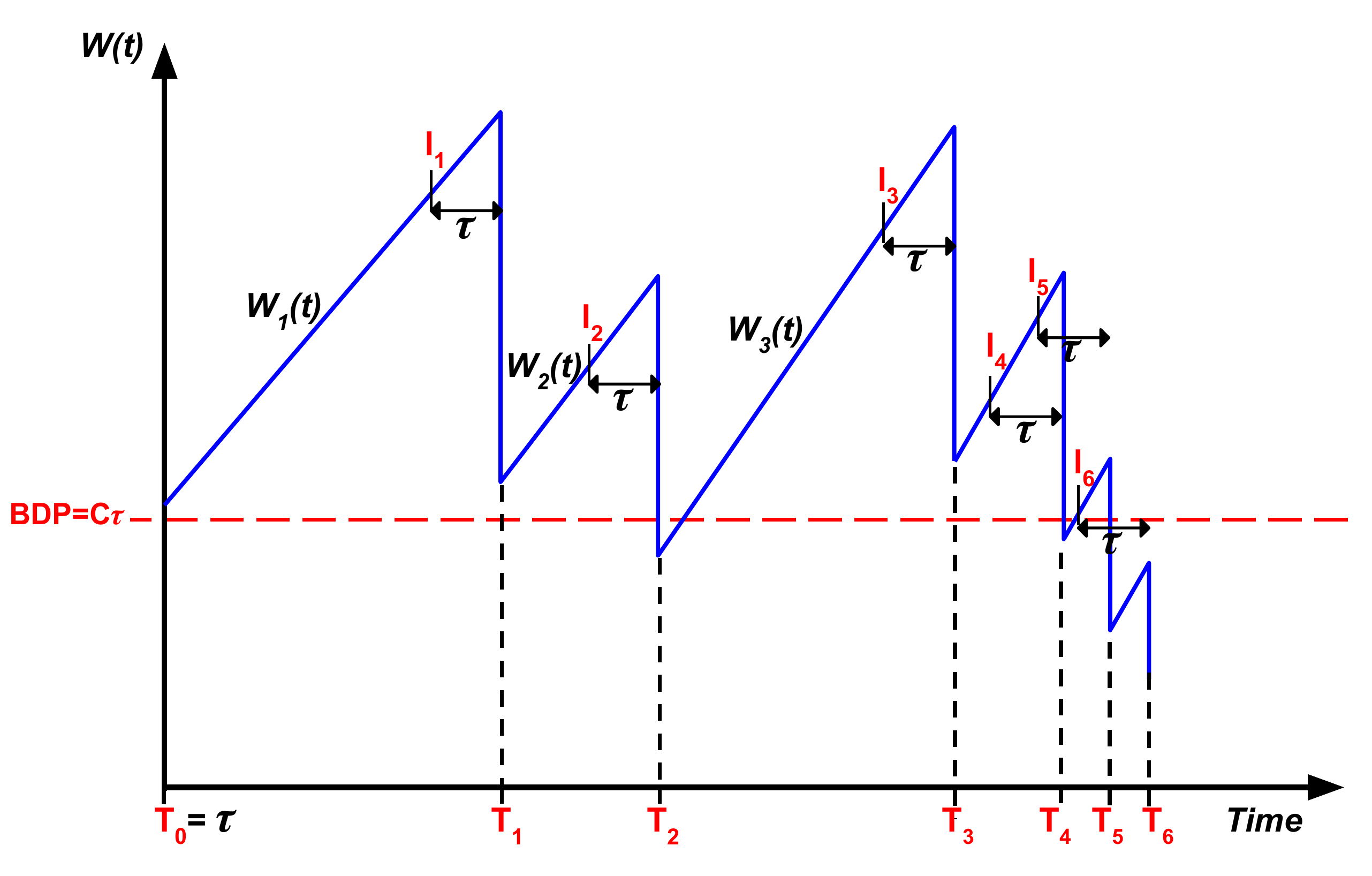}
\caption{Example trajectory of Reno's congestion window. $l_i$ is the time when loss occurs at the congestion point (\emph{e.g.} router). $T_i$ is the time of the $i$th loss indication.}
\label{fig:renofunc}
\end{figure*}

Note that $W(t)$ and $p(t)$ are viewed from the perspective of the congestion point (\emph{e.g.} a router) where the loss is being generated. Therefore, whenever a loss occurs, the subsequent reduction in the window size (multiplicative decrease) is not reflected in $W(t)$ until after a delay of approximately $\tau$ seconds. This is illustrated in Figure \ref{fig:renofunc}, which shows an example trajectory of the \cwnd function. Each time a loss $i$ occurs at time $l_i$ at a congestion point, a corresponding loss indication is reflected in $W(t)$ at time $T_i=l_i+\tau$.
The caveat of using (\ref{eq:solveForT}) to compute $T$ is that $W(t)$ may have changed sometime in the time interval $[t_0,t_0+T]$ (which can happen if a loss indication is scheduled in this interval; we call this a \textit{pending loss indication} (PLI)). In such a case, the solution is to project the current $W(t)$ until the next loss indication, update $W(t)$ to a new function, and use this new function to generate a new loss event. Once a new loss event is generated, the process may need to repeat until we either produce a loss event that takes place before the next PLI or until we run out of PLIs.

Another complication may arise with certain probability of loss models. For example, in this work we consider the following model:
\begin{align*}
p(t) &= \left(1-\frac{C\tau}{W(t)}\right)^+.
\end{align*}
As a consequence, $\lambda(t)=0$ whenever $W(t) < C\tau$. This is depicted in Figure \ref{fig:renofunc}, where losses only occur when $W(t) \geq C\tau$. In order to obtain an analytical solution for $T$ during the $i$th loss event, we can first compute $T_{BDP}$, the time at which $W(t)$ reaches $C\tau$, or the  bandwidth-delay product (BDP). Then, let
 $t_0=\max{(T_{BDP},l_i)}-T_{i-1}$, where 
 $T_{i-1}$ is the time of the most recent loss indication and  
 $l_i$ is the time of the most recent loss event at the congestion point.

Another feature of the simulation framework is the ability to generate multiple parallel flows. This feature is especially important for validating models that use a system of differential equations to characterize the behavior of congestion control algorithms. The output of such models (\emph{e.g.} \cwnd) usually describes the behavior of the average flow in a large population of flows. Indeed, in Section \ref{simulations}, we note that the average \cwnd size from simulation results matches closer to the steady-state value of the DE models as we increase the number of flows in the simulation. 

When multiple flows are involved, $T_{BDP}$ is the time at which the sum of their congestion windows reaches the BDP, and $l_i$ is the time at which the most recent loss (across all flows) occurred. We must compute $t_0$ for each flow, which is given by
\begin{align*}
t_{0,f}=\max{(T_{BDP},l_i)}-T_{i-1,f},
\end{align*}
where $T_{i-1,f}$ is the most recent loss indication of flow $f$.
$T$ is then computed using the following equation:
\begin{align}
\frac{1}{\tau}\sum_{f=1}^N \int_{t_{0,f}}^{t_{0,f}+T}W_f(t)p_f(t)dt = -\ln(u).
\label{eq:genmultflowloss}
\end{align}
Any time a new loss event is generated, we must also choose a flow that will suffer the loss. The flow is picked based on its congestion window size at the time the loss is scheduled to occur (flows with larger windows are more susceptible to suffer a loss).

\subsection{Pseudocode}
Loss generation can be described by the pseudocode in \texttt{GeneratePoiLoss}. This function is called from the main procedure each time a loss is occurring at the congestion point in a given interval. (So, for the example in Figure \ref{fig:renofunc}, \texttt{GeneratePoiLoss} would be called in the intervals containing the events $l_i$, $i\in\{1,\dots,6\}$.) The arguments of the function are as follows: $pendingLITs$ is a two-dimensional matrix whose first row is a list of pending loss indication times, and whose second row contains the corresponding flows that will suffer the losses. $LLIs$ is an array that keeps record of the last loss indication times of all flows. $GLLI$ is the most recent loss indication. $T_l$ is the time of the most recent loss event. $\wloss$ is an array containing the \cwnd sizes of all flows immediately before their most recent loss events. $p(t)$ is a probability of loss function and $\tau$ is the round-trip time.
 
\begin{algorithm}
\begin{algorithmic}
\Function{GeneratePoiLoss}{$pendingLITs$,  $LLIs$, $GLLI$, $T_l$, $\wloss$, $p(t)$, $\tau$}
\LineComment{$LLT$: last loss time at congestion point}
\LineComment{$pendingLITs$: a list of pending loss indication times and corresponding flows}
\State \textbf{Initialization:}
\State $GNPLI\gets pendingLITs.nextLossTime$ \Comment{next (global) pending loss indication time}
\State $LF\gets pendingLITs.nextFlow$ \Comment{the corresponding flow of the next loss event}
\State $T_{BDP}\gets$ time when sum of \cwnd's reaches BDP
\State $t_0\gets\max{(T_{BDP},T_l)}$
\State
\State $lossTime\gets \Call{computeT}{LLIs, GLLI, \wloss, t_0, \tau, p(t)}$
\State
\While{$lossTime \geq GNPLI$}
    \LineComment{next loss occurs after $GNPLI$, so:}
    \LineComment{(1) determine the duration of the current congestion epoch for flow $LF$:}
    \State $I\gets GNPLI-LLIs(f)$ 
    \LineComment{(2) the window function is changed at $GNPLI$, and we are looking at a new}
    \LineComment{congestion epoch, so update relevant variables}
    \State $\wloss \gets W_{LF}(I)$ \Comment{get the $\wloss$ value of the next congestion epoch for flow $LF$}
    \State $GLLI\gets GNPLI$
    \State $LLIs(LF)\gets GLLI$
    \If{$NPLI.isEmpty$}
    	\State $NPLI\gets \infty$
    \Else
	    \State $GNPLI\gets pendingLITs.nextLossTime$
	    \State $LF\gets pendingLITs.nextFlow$
    \EndIf
    \LineComment{(3) generate a new loss event at congestion point}
    \State Recompute $T_{BDP}$
    \State $t_0\gets\max({T_{BDP},GLLI})$
    \State $lossTime\gets \Call{computeT}{LLIs, GLLI, \wloss, t_0, \tau, p(t)}$
\EndWhile
\State
\LineComment{schedule the next loss indication event}
\State $pendingLITs.add(lossTime+\tau)$
\State \textbf{return} $(lossTime, pendingLITs)$
\EndFunction

\State
\Function{computeT}{$LLIs$, $GLLI$, $\wloss$, $t_0$, $\tau$, $p(t)$}
\State $u \gets rand()$ \Comment{generate a number from uniform distr.}
\State construct $W_f(t),\ \forall f\in\{1,\dots,N\}$ using $\wloss$
\State $t_{0,f}\gets t_0-LLIs(f),\ \forall f\in\{1,\dots,N\}$
\LineComment{to generate the next loss interval:}
\State Use Equation (\ref{eq:genmultflowloss}) to compute $T$, keep only real, positive roots
\State $lossTime\gets GLLI+t_0+T$
\EndFunction
\end{algorithmic}
\end{algorithm}

For the example in Figure \ref{fig:renofunc}, where there is only one flow, the procedure outlined in the pseudocode would do the following:
\begin{enumerate}
\item At time $t=0$, a loss occurred at the congestion point (not shown in the figure), so a pending loss indication was scheduled for $T_0=\tau$. 
\item Also at the time of the loss (at $t=0$), a new loss time was generated using \texttt{GeneratePoiLoss}. This loss time is $l_1$. Since $l_1$ occurs after the next pending loss indication (which is at $T_0$), the \texttt{while} loop in \texttt{GeneratePoiLoss} is triggered. We integrate the \cwnd function from $t=0$ to $T_0=\tau$, compute a new $\wloss$ (which is the size of the window right before $T_0$), and feed these values as parameters to \texttt{computeT}. The latter function computes the next loss arrival time; this is a new value of $l_1$. We compare this new $l_1$ to the next pending loss indication time (which in this case is $\infty$ since no other pending loss indications have been scheduled after $T_0$). Since $l_1 < \infty$, we exit the loop and have a new loss time of $l_1$ and pending loss indication $T_1=l_1+\tau$.
\item The main procedure iterates until it reaches the interval containing $l_1$, at which point \texttt{GeneratePoiLoss} is called. The latter function generates $l_2$, and since $l_2$ occurs after the next pending loss indication time (which is $T_1$), we re-generate $l_2$ using the same procedure as for $l_1$.
\item The main procedure continues until it reaches the interval containing $T_1$, at which point the loss indication is processed (the window is halved and there is a new $\wloss$).
\item Loss events $l_2$ and $l_3$ and pending loss indications $T_2$ and $T_3$ are processed similarly.
\item At loss event $l_4$, a new loss time $l_5$ is generated. Since it appears before $T_4$, we simply schedule a pending loss at $T_5$ (no need to go through the \texttt{while} loop in \texttt{GeneratePoiLoss} as we did for the other losses).
\item At loss event $l_5$, $l_6$ is generated, but it occurs after the pending loss indication at $T_4$, which has not been processed yet. Hence, the \texttt{while} loop is triggered.
\end{enumerate}
\section{Acknowledgment}
This work was supported by the US Department of Energy under Contract DE-AC02-06CH11357 and by the National Science Foundation under Grant No. CNS-1413998.

\bibliography{ref} 
\bibliographystyle{IEEEtran}
\end{document}